\pdfoutput=1
\documentclass[acmsmall]{acmart}

\usepackage[english]{babel}

\usepackage{amsfonts}
\usepackage{bbm}
\usepackage{bm}
\usepackage{mathtools}
\usepackage{scalerel} %
\usepackage{pifont}

\usepackage{booktabs}
\usepackage{subcaption}
\usepackage{adjustbox}
\usepackage{graphicx}
\usepackage{multirow}
\usepackage[justification=justified]{caption} %

\usepackage[inline]{enumitem}
\usepackage[capitalize]{cleveref}  %

\usepackage{tikz}
\usetikzlibrary{arrows.meta}
\usetikzlibrary{backgrounds}
\usetikzlibrary{positioning, fit, mindmap, trees, calc, tikzmark, shapes}
\usetikzlibrary{shapes.arrows, fadings, automata, tikzmark, decorations.pathreplacing, patterns}
\usetikzlibrary{arrows.meta}
\usetikzlibrary{intersections}
\usepackage{tkz-euclide}
\usepackage{tikzpeople}

\usepackage{csvsimple}
\usepackage{pgfplots}
\usepgfplotslibrary{groupplots}
\pgfplotsset{compat=1.14}

\usepackage{sistyle}
\SIthousandsep{,}

\usepackage{soul}
\setstcolor{red}

\usepackage{xcolor}

\usepackage{balance}

\definecolor{tea_green}{RGB}{214, 234, 193}
\definecolor{hint_green}{RGB}{226,246,209}
\definecolor{Madang}{RGB}{190,235,159}
\definecolor{yellow_green}{RGB}{198,222,119}
\definecolor{link_water}{RGB}{221, 232, 250}
\definecolor{celestial_blue}{RGB}{52, 152, 219}
\definecolor{shakespeare}{RGB}{85, 154, 193}
\definecolor{buttermilk}{RGB}{255,242,174}
\definecolor{chardonnay}{RGB}{250,196,114}
\definecolor{rajah}{RGB}{253,180,98}
\definecolor{fog}{RGB}{213, 193, 234}
\definecolor{melon}{RGB}{254,191,181}
\definecolor{sundown}{RGB}{249, 180, 181}
\definecolor{mona_lisa}{RGB}{246,152,134}
\definecolor{salmon}{RGB}{242,131,107}

\definecolor{saltpan}{RGB}{238, 243, 232}
\definecolor{aqua_spring}{RGB}{232, 243, 232}
\definecolor{tea_green}{RGB}{214, 234, 193}
\definecolor{Madang}{RGB}{190,235,159}
\definecolor{fringy_flower}{RGB}{194, 234, 193}
\definecolor{aero_blue}{RGB}{193, 234, 213}
\definecolor{pixie_green}{RGB}{183,214,170}
\definecolor{french_pass}{RGB}{195,232,246}
\definecolor{ice_cold}{RGB}{169,232,220}
\definecolor{pale_turquoise}{RGB}{172,240,242}
\definecolor{cruise}{RGB}{179,226,205}
\definecolor{sail}{RGB}{163,205,235}
\definecolor{spindle}{RGB}{179,205,227}
\definecolor{link_water}{RGB}{221, 232, 250}
\definecolor{periwinkle}{RGB}{203,213,232}
\definecolor{zanah}{RGB}{220, 233, 213}
\definecolor{frostee}{RGB}{217, 231, 214}
\definecolor{opal}{RGB}{199, 221, 211}
\definecolor{jet_stream}{RGB}{188, 214, 210}
\definecolor{skeptic}{RGB}{153, 187, 167}
\definecolor{hint_green}{RGB}{226,246,209}
\definecolor{snow_flurry}{RGB}{230,245,201}
\definecolor{surf_crest}{RGB}{205,230,208}
\definecolor{yellow_green}{RGB}{198,222,119}
\definecolor{cream}{RGB}{255,255,204}
\definecolor{pale_prim}{RGB}{255,255,179}
\definecolor{spring_sun}{RGB}{242,243,195}
\definecolor{portafino}{RGB}{245,237,160}
\definecolor{buttermilk}{RGB}{255,242,174}
\definecolor{cream_brulee}{RGB}{255, 229, 151}
\definecolor{dairy_cream}{RGB}{254,226,189}
\definecolor{champagne}{RGB}{254,217,166}
\definecolor{chardonnay}{RGB}{250,196,114}
\definecolor{manhattan}{RGB}{226,180,125}
\definecolor{rajah}{RGB}{253,180,98}
\definecolor{early_dawn}{RGB}{252,243,218}
\definecolor{egg_shell}{RGB}{238, 234, 215}
\definecolor{selago}{RGB}{243, 232, 243}
\definecolor{quartz}{RGB}{219,223,238}
\definecolor{fog}{RGB}{213, 193, 234}
\definecolor{languid_lavender}{RGB}{222,203,228}
\definecolor{watusi}{RGB}{254,221,207}
\definecolor{coral_andy}{RGB}{243,204,205}
\definecolor{cosmos}{RGB}{248,209,210}
\definecolor{melon}{RGB}{254,191,181}
\definecolor{azalea}{RGB}{234, 193, 194}
\definecolor{beauty_bush}{RGB}{235, 185, 179}
\definecolor{sundown}{RGB}{249, 180, 181}
\definecolor{mona_lisa}{RGB}{246,152,134}
\definecolor{salmon}{RGB}{242,131,107}

\definecolor{summer_sky}{RGB}{58, 151, 233}
\definecolor{chateau_green}{RGB}{72, 179, 96}
\definecolor{matisse}{RGB}{25, 104, 167}
\definecolor{allports}{RGB}{31, 106, 125}
\definecolor{sun_shade}{RGB}{255, 144, 68}
\definecolor{flamingo}{RGB}{237, 88, 85}
\definecolor{studio}{RGB}{128, 91, 160}

\definecolor{maya_blue}{RGB}{102, 204, 255}
\definecolor{feijoa}{RGB}{178,223,138}
\definecolor{sushi}{RGB}{117, 168, 47}
\definecolor{norway}{RGB}{158, 194, 132}
\definecolor{japanese_laurel}{RGB}{53, 116, 40}
\definecolor{see_green}{RGB}{161,228,195}
\definecolor{monte_carlo}{RGB}{135,204,194}
\definecolor{granny_smith_apple}{RGB}{150,214,150}
\definecolor{moss_green}{RGB}{170,216,176}
\definecolor{chateau_green}{RGB}{72, 179, 96}
\definecolor{opal}{RGB}{164,207,190}
\definecolor{acapulco}{RGB}{117, 170, 148}
\definecolor{viridian}{RGB}{55, 137, 122}
\definecolor{amazon}{RGB}{56, 123, 84}
\definecolor{asparagus}{RGB}{123, 160, 91}
\definecolor{fruit_salad}{RGB}{91, 160, 94}
\definecolor{puerto_rico}{RGB}{72, 179, 150}
\definecolor{mountain_meadow}{RGB}{0, 163, 136}
\definecolor{matisse}{RGB}{25, 104, 167}
\definecolor{allports}{RGB}{31, 106, 125}
\definecolor{astral}{RGB}{55, 111, 137}
\definecolor{spring_leaves}{RGB}{46, 83, 117}
\definecolor{biscay}{RGB}{44, 62, 80}
\definecolor{midnight}{RGB}{0, 29, 50}
\definecolor{amethyst}{RGB}{153, 102, 204}
\definecolor{studio}{RGB}{128, 91, 160}
\definecolor{tapestry}{RGB}{194, 109, 132}
\definecolor{atomic_tangerine}{RGB}{255, 153, 102}
\definecolor{amber}{RGB}{255, 191, 0}
\definecolor{casablanca}{RGB}{244, 178, 84}
\definecolor{california}{RGB}{233, 140, 58}
\definecolor{tomato}{RGB}{255, 97, 56} 
\definecolor{alizarin}{RGB}{233, 58, 64}

\definecolor{linen}{RGB}{251, 239, 227}
\definecolor{double_pearl_lusta}{RGB}{253, 242, 208}
\definecolor{oasis}{RGB}{253, 242, 208}
\definecolor{milan}{RGB}{255, 254, 169}
\definecolor{texas}{RGB}{245, 232, 123}
\definecolor{maize}{RGB}{249, 212, 156}

\definecolor{turmeric}{RGB}{211, 178, 76}
\definecolor{saffron}{RGB}{249,193,62}
\definecolor{my_sin}{RGB}{255, 176, 59}
\definecolor{tree_poppy}{RGB}{246, 154, 27}
\definecolor{jaffa}{RGB}{240, 131, 58}
\definecolor{crusta}{RGB}{254, 127, 44}
\definecolor{tahiti_gold}{RGB}{223, 102, 36}
\definecolor{outrageous_orange}{RGB}{255, 100, 45}
\definecolor{safety_orange}{RGB}{254, 106, 0}

\definecolor{azalea}{RGB}{251, 196, 196}
\definecolor{oyster_pink}{RGB}{238,206,205} 
\definecolor{coral_candy}{RGB}{242,208,205} 
\definecolor{baby_pink}{RGB}{246, 194, 192}
\definecolor{petite_orchid}{RGB}{223, 157, 155}
\definecolor{apricot}{RGB}{241,140,122}
\definecolor{NY_pink}{RGB}{228,136,113}
\definecolor{carmine_pink}{RGB}{231, 76, 60}
\definecolor{deep_carmine_pink}{RGB}{236, 50, 67}

\definecolor{wewak}{RGB}{244, 143, 150}
\definecolor{light_coral}{RGB}{244, 127, 123}
\definecolor{bittersweet}{RGB}{255,111,105}
\definecolor{carnation}{RGB}{245, 80, 86}
\definecolor{flamingo}{RGB}{237, 88, 85}
\definecolor{sunset_orange}{RGB}{242,89,75}
\definecolor{ku_crimson}{RGB}{243, 0, 25}
\definecolor{amaranth}{RGB}{234,46,73}
\definecolor{valencia}{RGB}{214, 87, 70}
\definecolor{chilean_firegongs }{RGB}{215, 87, 44}
\definecolor{mexican_red}{RGB}{170, 41, 37}

\definecolor{napa}{RGB}{163, 154, 137}

\definecolor{athens_gray}{RGB}{236, 240, 241}
\definecolor{gallery}{RGB}{240,240,240}
\definecolor{mercury}{RGB}{230,230,230}
\definecolor{platinum}{RGB}{228,228,228}
\definecolor{silver}{RGB}{191,191,191}
\definecolor{aluminum}{RGB}{153,153,153}
\definecolor{ship_gray}{RGB}{77,77,77}
\definecolor{tuatara}{RGB}{67, 67, 67}

\definecolor{malibu}{RGB}{110, 180, 240}
\definecolor{celestial_blue}{RGB}{52, 152, 219}
\definecolor{curious_blue}{RGB}{41, 128, 185}
\definecolor{french_blue}{RGB}{0, 112, 182}
\definecolor{matisse}{RGB}{25, 104, 167}
\definecolor{shakespeare}{RGB}{85, 154, 193}
\definecolor{seagull}{RGB}{128,177,211}
\definecolor{jelly_bean}{RGB}{45, 126, 150}
\definecolor{venice_blue}{RGB}{87, 135, 105}
\definecolor{boston_blue}{RGB}{68, 147, 161}

\definecolor{turquoise}{RGB}{41,217,194}
\definecolor{java}{RGB}{2,190,196}
\definecolor{riptide}{RGB}{141,211,199}
\definecolor{mountain_meadow}{RGB}{0, 163, 136}
\definecolor{free_speech_aquamarine}{RGB}{0, 156, 114}

\definecolor{cosmic_latte}{RGB}{222, 247, 229}
\definecolor{chinook}{RGB}{163, 232, 178}
\definecolor{padua}{RGB}{121, 189, 143}
\definecolor{ocean_green}{RGB}{79, 176, 112}
\definecolor{pastel_green}{RGB}{107, 227, 135}
\definecolor{chateau_green}{RGB}{69, 191, 85}
\definecolor{RoyalBlue}{RGB}{69, 191, 85}
\definecolor{pigment_green}{RGB}{0, 175, 79}
\definecolor{fern}{RGB}{101,197,117}
\definecolor{killarney}{RGB}{56, 113, 66}

\newtheorem{assumption}{Assumption}

\DeclareMathOperator*{\argmax}{arg\,max}

\newcommand{\si}[1]{{\scriptstyle \mathcal{S}_i#1}}
\newcommand{\di}[1]{{\scriptstyle \mathcal{D}_i#1}}

\newcommand{\czq}[1]{\textcolor{black}{{#1}}}

\renewcommand{\arraystretch}{1.2}

\AtBeginDocument{%
  \providecommand\BibTeX{{%
    \normalfont B\kern-0.5em{\scshape i\kern-0.25em b}\kern-0.8em\TeX}}}

\setcopyright{acmcopyright}
\copyrightyear{2022}
\acmYear{2022}
\acmDOI{XXXXXXX.XXXXXXX}

\acmJournal{TOIS}
\acmVolume{37}
\acmNumber{4}
\acmArticle{111}
\acmMonth{8}

\begin{document}

\title{Studying the Impact of Data Disclosure Mechanism in Recommender Systems via Simulation}

\author{Ziqian Chen}
\email{eric.czq@alibaba-inc.com}
\affiliation{%
  \department{Damo Academy}
  \institution{Alibaba Group}
  \city{Hangzhou}
  \state{Zhejiang}
  \postcode{311121}
  \country{China}
}

\author{Fei Sun}
\authornote{Fei Sun is the corresponding author and now works at ICT, CAS.}
\orcid{0000-0002-6146-148X}
\email{ofey.sf@alibaba-inc.com}
\affiliation{%
  \department{Damo Academy}
  \institution{Alibaba Group}
  \city{Beijing}
  \postcode{100102}
  \country{China}
}

\author{Yifan Tang}
\email{yifan.tang95@gmail.com}
\affiliation{%
  \department{Luohan Academy}
  \institution{Alibaba Group}
  \city{Hangzhou}
  \state{Zhejiang}
  \postcode{311121}
  \country{China}
}

\author{Haokun Chen}
\email{hankel.chk@alibaba-inc.com}
\affiliation{%
  \department{Damo Academy}
  \institution{Alibaba Group}
  \city{Hangzhou}
  \state{Zhejiang}
  \postcode{311121}
  \country{China}
}

\author{Jinyang Gao}
\email{jinyang.gjy@alibaba-inc.com}
\affiliation{%
  \department{Damo Academy}
  \institution{Alibaba Group}
  \city{Hangzhou}
  \state{Zhejiang}
  \postcode{311121}
  \country{China}
}

\author{Bolin Ding}
\email{bolin.ding@alibaba-inc.com}
\affiliation{%
  \department{Damo Academy}
  \institution{Alibaba Group}
  \city{Seattle}
  \state{WA}
  \postcode{98004}
  \country{United States}
}

\begin{abstract}

Recently, privacy issues in web services that rely on users' personal data have raised great attention.
Despite that recent regulations force companies to offer choices for each user to opt-in or opt-out of data disclosure, real-world applications usually only provide an ``all or nothing'' binary option for users to either disclose all their data or preserve all data with the cost of no personalized service.

In this paper, we argue that such a binary mechanism is not optimal for both consumers and platforms.
To study how different privacy mechanisms affect users' decisions on information disclosure and how users' decisions affect the platform's revenue, we propose a privacy aware recommendation framework that gives users fine control over their data. 
In this new framework, users can proactively control which data to disclose based on the trade-off between anticipated privacy risks and potential utilities.
Then we study the impact of different data disclosure mechanisms via simulation with reinforcement learning due to the high cost of real-world experiments.
The results show that the platform mechanisms with finer split granularity and more unrestrained disclosure strategy can bring better results for both consumers and platforms than the ``all or nothing''  mechanism adopted by most real-world applications.

\end{abstract}



\maketitle

\section{Introduction}

Recommender systems play an essential role on today's web service platforms, e.g., e-commerce~\cite{Linden:IC03:Amazon,xie21explore} and social media~\cite{Covington:recsys16:Deep,Ying:kdd18:Graph}, since they can reduce users' cognitive load by automatically offering personalized services that match their interests and needs \cite{chin2007information}.
While the recommender systems greatly facilitate the distribution and acquisition of information, they also bring critical privacy concerns due to unsolicited gathering users' demographical and behavioral data ~\cite{Smith:mis96:Information,zhang2021membership}. %
Several regulations have been proposed recently to better protect personal data, e.g., General Data Protection Regulation (GDPR) in the European Union and the California Privacy Rights Act (CPRA) in the United States. %

On the one hand, various privacy-preserving methods have been proposed to protect users' data from leakage or abuse, like federated learning~\cite{Lin:sigir20:Meta,Muhammad:kdd20:FedFast,Qi:emnlp20:Privacy,Minto:recsys21:Stronger} and differential privacy (DP)~\cite{McSherry:kdd09:Differentially,Berlioz:recsys15:Applying,shin2018privacy,Gao:sigir20:DPLCF}.
On the other hand, to comply with laws like GDPR and CCPA, most platforms now provide users the choice to opt-in (under GDPR) or opt-out (under CCPA) of data disclosure.
Previous studies have shown that giving users control over their data closure can reduce their privacy concerns~\cite{Zhang2014-oa,Chen:CHI18:This}.
However, in practice, these platforms mostly only provide a very coarse-grained option for each user, named ``all or nothing'' binary mechanism in this paper, i.e., disclosing all data or none at all.
Usually, if the users choose to not disclose their data, they will either not be able to continue using the applications \footnote{Some applications will refuse to serve the users who refuse to disclose data by turning off the software when users clicks the refusal button, so as to force users to approve their data disclosure disclaimer and collect their data.} or not be able to enjoy the precisely personalized services.

This raises a question, \textit{is such an ``all or nothing'' binary mechanism the optimal choice?}
Obviously, some privacy sensitive users might choose not to disclose their data.
In this case, these users cannot
enjoy the benefits of personalized services.
At the same time, the platform revenues from these privacy sensitive users will decrease due to the disappointed personalized services and the platform also loses their data to train a better model.
It seems that such a strict mechanism might not be a satisfying choice for both parties in the ecology.
Therefore, we wonder whether there exists a better mechanism to take both the gains of the users and the revenues of the platform into account?

This paper aims to study how different privacy mechanisms affect users' decisions on information disclosure and how their decisions affect the recommendation model's performance and the platform's revenue.
For this purpose, we propose a privacy aware recommendation framework under \textit{privacy calculus theory}~\cite{Laufer:si77:Privacy,Culnan:os99:Information}.
Under this new setting, users need to calculate the trade-off between the anticipated privacy risks and the potential utilities, then \textit{proactively control which data to disclose}. 
In this way, all users' dispersed privacy preferences are fully accommodated.

For service providers, they naturally want to entice users to disclose as much data as possible.
For end users, they want to figure out how to enjoy the benefits of personalized services with minimal privacy risks.
Formally speaking, under this privacy aware recommendation task setting, we aim to study \textit{what will happen if the platforms give users fine-grained control over their personal data}.
More specifically, we investigate questions including:
\begin {enumerate} [itemsep=5pt, topsep=6pt, label=\roman*\upshape)] %
\item How do different platform mechanisms affect users' decisions in information disclosure? Is the ``all or nothing'' binary mechanism the best choice for the platform?
\item How do different recommendation models affect users' decisions in information disclosure? Can a platform attract users to disclose more data by optimizing the model to provide better services?
\end {enumerate}

To answer these questions, we first formulate our idea in formal settings. %
Following current researches in economics~\cite{lin2019valuing,tang2019value}, we model the privacy cost as a linear summation of the user's disclosed personal data, meaning that the user loses control over such disclosed data, which also fits a fundamental notion in privacy calculus, i.e., the control over the data. %
Then recommendation performance (e.g., NDCG) is employed as the potential utility from users' disclosed data.
To formally define user privacy decisions, we formulate the platform mechanisms using two components, i.e., data split rule and data disclosure choice space, which define the choices a user can take.
Based on these simplified settings, we now can conduct experiments with different platform mechanisms or recommendation models to find answers to the above questions.

However, there is one big challenge for directly realizing our idea in real-world applications.
Direct deployment of the proposed framework in real-world applications might seriously harm the end users’ experiences and the revenues of platforms.
To address this challenge, inspired by the success of simulation studies in recommender systems~\cite{Ie:arxiv19:RecSim,krauth2020offline,lucherini2021t,yao2021measuring}, we propose to use simulations to study the effects of the proposed framework.
Specifically, we propose a reinforcement learning method to simulate users' privacy decision making on two benchmark datasets with three representative recommendation models and three user types (i.e. different privacy sensitivity).
The experimental results show that the platform mechanism with finer split granularity and more unconstrained disclosure strategy can bring better results for both end users and platforms than ``all or nothing'' binary mechanism adopted by most platforms.
In addition to mechanism design, the results also point out that optimizing model is another option for the platform to collect more data while protecting user privacy.

Our main contributions can be summarized as following:
\begin{itemize} [itemsep=5pt, topsep=10pt,] %
    \item We study an important and new problem in recommender systems and privacy protection, the effects of platform mechanisms on users' privacy decision.
    \item We propose a privacy aware recommendation framework that gives users control over their personal data. To the best of our knowledge, this is the first work to give users fine-grained control over implicit feedback data in recommendation.
    \item We formulate the process of users' privacy decision making and the platform's data disclosure mechanisms using mathematics language. Then we instantiate the platform's mechanisms with one data split rule and three data disclosure strategies that we proposed. %
    \item We propose a reinforcement learning method to simulate users' privacy decision making. The extensive simulations are conducted on two benchmark datasets with three representative recommendation models.
    \item %
    The extensive experimental results show the effectiveness of our proposed framework in protecting users' privacy.
    The results also shed some light on data disclosure mechanism design and model optimization.
\end{itemize}

\section{Framework Formulation}

\subsection{Overview} %
\tikzset{
  FARROW/.style={arrows={-{Latex[length=1.25mm, width=1.mm]}}},
  DFARROW/.style={arrows={{Latex[length=1.25mm, width=1.mm]}-{Latex[length=1.25mm, width=1.mm]}}},
  behavior/.style = {circle, fill=monte_carlo, minimum width=1.2em, align=center, inner sep=0, outer sep=0, font=\tiny},
  feature/.style = {circle, fill=salmon, minimum width=1.2em, align=center, inner sep=0, outer sep=0, font=\tiny},
  encoder/.style = {rectangle, fill=Madang!82, minimum width=6em, minimum height=2em, align=center, rounded corners=3},
  emb_layer/.style = {rectangle, fill=languid_lavender!72, minimum width=11em, minimum height=2em, align=center, rounded corners=3},
  project/.style = {rectangle, fill=hint_green, minimum width=7em, minimum height=2.4em, align=center, rounded corners=2},
  ds/.style={
       rectangle split,
       rectangle split part align=base,
       rectangle split horizontal=true,
       rectangle split draw splits=true,
       rectangle split parts=5,
       rectangle split part fill={athens_gray!80, athens_gray!80, athens_gray!80, athens_gray!80, athens_gray!80},
       draw=black, %
       very thin,
       minimum height=1.2em,
       minimum width=2em,
       text width=0.4em,
       inner sep=.5pt,
       text centered,
       font=\tiny,
       text=gray,
       },
}

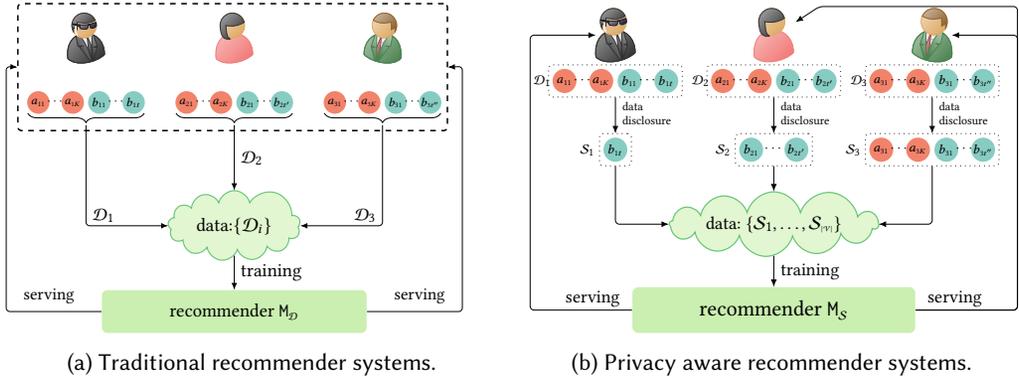
\begin{figure}
\centering
\begin{subfigure}[b]{0.48\textwidth}
\resizebox{0.92\linewidth}{!}{
\begin{tikzpicture}
    \node[] (to) at (0, 0) {};
    
    \node[person, female, hair=black, shirt=flamingo, skin=skin, minimum size=0.7cm, below of=to, node distance=1cm] (t_ub) {};
    \node[businessman, mirrored, minimum size=0.7cm, right of=t_ub, node distance=2.8cm] (t_uc) {};
    \node[maninblack, minimum size=0.7cm, left of=t_ub, node distance=2.8cm] (t_ua) {};
    
    \foreach \x/\y in {a/1, b/2, c/3}
    {
    \node [feature, below of=t_u\x, node distance=1.2cm, xshift=-0.9cm] (f_t\x1) {$a_{\scaleto{\y1}{2.5pt}}$};
    \node [right of=f_t\x1, node distance=0.325cm, font=\tiny, align=center] (f_t\x2) {$\cdots$};
     \node [feature, right of=f_t\x2, node distance=0.325cm] (f_t\x3) {$a_{\scaleto{\y K}{2.4pt}}$};
     \node [behavior, right of=f_t\x3, node distance=0.5cm, font=\tiny] (b_t\x1) {$b_{\scaleto{\y1}{2.5pt}}$};
     \node [right of=b_t\x1, node distance=0.325cm, font=\tiny] (b_t\x2) {$\cdots$};
    }
    
    \node [behavior, right of=b_ta2, node distance=0.325cm] (b_ta3) {$b_{\scaleto{1t}{2.5pt}}$};
    \node [behavior, right of=b_tb2, node distance=0.325cm] (b_tb3) {$b_{\scaleto{2t'}{3.2pt}}$};
    \node [behavior, right of=b_tc2, node distance=0.325cm] (b_tc3) {$b_{\scaleto{3t''}{3pt}}$};
     
     \foreach \x in {a, b, c}
     {
     \draw [decorate, decoration={mirror, brace, amplitude=5pt}] ([xshift=-0.5mm, yshift=-0.5mm] f_t\x1.south west) -- ([xshift=0.5mm, yshift=-0.5mm] b_t\x3.south east) node[midway,yshift=0.5cm] (d_t\x) {};
     }
     
     \node [cloud, cloud puffs=12.8, cloud ignores aspect, draw=fern, semithick, fill=hint_green, minimum width=25mm, align=center, inner ysep=4.5pt, below of=to, node distance=4.5cm] (data_t) {data:$\{\mathcal{D}_i\}$};
     
     \node[encoder, below of=data_t, node distance=1.6cm, minimum width=5cm, minimum height=0.8cm,] (rec_model) {recommender $\texttt{M}_{{\scriptscriptstyle \mathcal{D}}}$};
     
     \node[dashed, thick, draw=black, fit={(t_ua) (t_uc) ([yshift=-2mm] f_ta1) ([yshift=-2mm] b_tc3.south east)}, inner sep=5, rounded corners=2] (t_user_side) {};
     
     \draw[FARROW, rounded corners=2] ([yshift=-6mm] d_ta.south) |- (data_t.west) node [pos=0.5, yshift=2mm, right, black, align=left, font=\small ]  { $\mathcal{D}_1$};
     \draw[FARROW, rounded corners=2] ([yshift=-6mm] d_tc.south) |- (data_t.east) node [pos=0.5, yshift=2mm, left, black, align=left, font=\small ]  {$\mathcal{D}_3$};
     \draw[FARROW] ([yshift=-6mm] d_tb.south) -> (data_t.north) node [pos=0.5, right, black, align=left, font=\small ]  {$\mathcal{D}_2$};
     
     \draw[FARROW] ( data_t.south) -> (rec_model.north)  node [pos=0.5, right, black, align=left] (box_c_a){training};
     \draw[FARROW, rounded corners=2] (rec_model.west) -- ++(-1.8, 0) -- ++(0, 4.57) -- (t_user_side.west);  
     \draw[FARROW, rounded corners=2] (rec_model.east) -- ++(1.8, 0) -- ++(0, 4.57) -- (t_user_side.east);  
     \node[left of=rec_model, node distance=3.5cm, yshift=3mm, font=\small] {serving};
     \node[right of=rec_model, node distance=3.5cm, yshift=3mm, font=\small] {serving};
     
\end{tikzpicture}}
\caption{Traditional recommender systems.}
\label{fig:traditional_rec}
\end{subfigure}
\hfil
\begin{subfigure}[b]{0.48\textwidth}
\resizebox{\linewidth}{!}{
    \begin{tikzpicture}

    \node[person, female, hair=black, shirt=flamingo, skin=skin, minimum size=0.7cm] (ub) at (0,0) {};
    \node[businessman, mirrored, minimum size=0.7cm, right of=ub, node distance=2.8cm] (uc) {};
    \node[maninblack, minimum size=0.7cm, left of=ub, node distance=2.8cm] (ua) {};

    \node [feature, below of=ua, node distance=0.8cm, xshift=-0.9cm, ] (f_a1) {$a_{\scaleto{11}{2.5pt}}$};
    \node [right of=f_a1, node distance=0.325cm, font=\tiny, align=center] (f_a2) {$\cdots$};
    \node [feature, right of=f_a2, node distance=0.325cm, ] (f_a3) {$a_{\scaleto{1K}{2.4pt}}$};
    \node [behavior, right of=f_a3, node distance=0.5cm, ] (b_a1) {$b_{\scaleto{11}{2.5pt}}$};
    \node [right of=b_a1, node distance=0.325cm, font=\tiny] (b_a2) {$\cdots$};
    \node [behavior, right of=b_a2, node distance=0.325cm] (b_a3) {$b_{\scaleto{1t}{2.5pt}}$};
    
    \node [left of=f_a1, node distance=0.4cm, font=\tiny] (d1) {$\mathcal{D}_1$};
    
    \node [behavior, below of=ua, node distance=2cm] (s_b_a3) {$b_{{\scaleto{1t}{2.5pt}}}$};

    \node[dotted, draw=black, fit={(f_a1) (b_a3) }, inner sep=2, rounded corners=2] (d_box_a) {};
    \node[dotted, draw=black, fit={(s_b_a3) }, inner sep=2, rounded corners=2] (s_box_a) {};
    \draw[FARROW] (d_box_a) -> (s_box_a) node [pos=0.45, right, black, font=\tiny, align=left]  {data\\ disclosure};
    
    \node [feature, below of=uc, node distance=0.8cm, xshift=-0.9cm] (f_c1) {$a_{\scaleto{31}{2.5pt}}$};
    \node [right of=f_c1, node distance=0.325cm, font=\tiny, align=center] (f_c2) {$\cdots$};
    \node [feature, right of=f_c2, node distance=0.325cm] (f_c3) {$a_{\scaleto{3K}{2.5pt}}$};
    \node [behavior, right of=f_c3, node distance=0.5cm] (b_c1) {$b_{\scaleto{31}{2.5pt}}$};
    \node [right of=b_c1, node distance=0.325cm, font=\tiny] (b_c2) {$\cdots$};
    \node [behavior, right of=b_c2, node distance=0.325cm] (b_c3) {$b_{\scaleto{3t''}{3pt}}$};
    \node [left of=f_c1, node distance=0.4cm, font=\tiny] (d3) {$\mathcal{D}_3$};
    
    \node [feature, below of=f_c1, node distance=1.2cm] (s_f_c1) {$a_{\scaleto{31}{2.5pt}}$};
    \node [below of=f_c2, node distance=1.2cm, font=\tiny, align=center] (s_f_c2) {$\cdots$};
    \node [feature, below of=f_c3, node distance=1.2cm] (s_f_c3) {$a_{\scaleto{3K}{2.5pt}}$};
    \node [behavior, below of=b_c1, node distance=1.2cm] (s_b_c1) {$b_{\scaleto{31}{2.5pt}}$};
    \node [below of=b_c2, node distance=1.2cm, font=\tiny] (s_b_c2) {$\cdots$};
    \node [behavior, below of=b_c3, node distance=1.2cm] (s_b_c3) {$b_{\scaleto{3t''}{3pt}}$};
    
    \node[dotted, draw=black, fit={(f_c1) (b_c3) }, inner sep=2, rounded corners=2] (d_box_c) {};
    \node[dotted, draw=black, fit={(s_f_c1) (s_b_c3) }, inner sep=2, rounded corners=2] (s_box_c) {};
    \draw[FARROW] (d_box_c) -> (s_box_c) node [pos=0.45, right, black, font=\tiny, align=left]  {data\\ disclosure};

    \node [feature, below of=ub, node distance=0.8cm, xshift=-0.9cm, ] (f_b1) {$a_{\scaleto{21}{2.5pt}}$};
    \node [right of=f_b1, node distance=0.325cm, font=\tiny, align=center] (f_b2) {$\cdots$};
    \node [feature, right of=f_b2, node distance=0.325cm,] (f_b3) {$a_{\scaleto{2K}{2.5pt}}$};
    \node [behavior, right of=f_b3, node distance=0.5cm, ] (b_b1) {$b_{\scaleto{21}{2.5pt}}$};
    \node [right of=b_b1, node distance=0.325cm, font=\tiny] (b_b2) {$\cdots$};
    \node [behavior, right of=b_b2, node distance=0.325cm] (b_b3) {$b_{\scaleto{2t'}{3.2pt}}$};
    \node [left of=f_b1, node distance=0.4cm, font=\tiny] (d2) {$\mathcal{D}_2$};
    
    \node [below of=ub, node distance=2cm, font=\tiny, align=center] (s_b_b2) {$\cdots$};
    \node [behavior, left of=s_b_b2, node distance=0.4cm,] (s_b_b1) {$b_{\scaleto{21}{2.5pt}}$};
    \node [behavior, right of=s_b_b2, node distance=0.4cm,] (s_b_b3) {$b_{\scaleto{2t'}{3.2pt}}$};
    
     \node[dotted, draw=black, fit={(f_b1) (b_b3) }, inner sep=2, rounded corners=2] (d_box_b) {};
    \node[dotted, draw=black, fit={(s_b_b1) (s_b_b3) }, inner sep=2, rounded corners=2] (s_box_b) {};
    \draw[FARROW] (d_box_b) -> (s_box_b) node [pos=0.45, right, black, font=\tiny, align=left]  {data\\ disclosure};
    
    \node [cloud, cloud puffs=12.3, cloud ignores aspect, draw=fern, semithick, fill=hint_green, minimum width=25mm, align=center, inner ysep=0pt, below of=ub, node distance=3.3cm, font=\small] (data_p) {data: $\{\mathcal{S}_1, \dots, \mathcal{S}_{\scaleto{|\mathcal{V}|}{3.2pt}}\}$};
    
    \node[left of = s_box_a, node distance=0.5cm, font=\tiny] {$\mathcal{S}_1$};
    \node[left of = s_b_b1, node distance=5mm, font=\tiny]  {$\mathcal{S}_2$};
    \node[left of = s_f_c1, node distance=5mm, font=\tiny]  {$\mathcal{S}_3$};
    
    \draw[FARROW, rounded corners=2] (s_box_a.south) |- (data_p.west);
    \draw[FARROW, rounded corners=2] ( s_box_c.south) |- (data_p.east);
    \draw[FARROW] (s_box_b.south) -> (data_p.north);
    
    \node[encoder, below of=ub, node distance=4.8cm, minimum width=5cm, minimum height=0.8cm, font=\small] (model_p) {\large recommender $\texttt{M}_{{\scriptscriptstyle \mathcal{S}}}$};
    
    \draw[FARROW] (data_p.south) -> (model_p.north) node [pos=0.5, right, black, align=left, font=\small] (box_c_a){training};
    
    \draw[FARROW, rounded corners=2] (model_p.west) -- ++(-1.8, 0) -- ++(0, 4.8) -- (ua.west); 
    \draw[FARROW, rounded corners=2] (model_p.east) -- ++(1.8, 0) -- ++(0, 4.8) -- (uc.east); 
    \draw[FARROW, rounded corners=2] (model_p.east) -- ++(1.8, 0) -- ++(0, 5.3) -- ++(-3.5, 0) -- (ub); 
    
    \node[left of=model_p, node distance=3.2cm, yshift=2mm, font=\small] {serving};
    \node[right of=model_p, node distance=3.2cm, yshift=2mm, font=\small] {serving};

    \end{tikzpicture} }
    \caption{Privacy aware recommender systems.}
    \label{fig:privacy_rec}
    \end{subfigure}
    \caption{Illustrative examples for two different recommender system frameworks.}
    \label{fig:framework}
\end{figure}

To study how different platform mechanisms affect users' decisions in information disclosure, we first need to incorporate the users' data disclosure decisions into the recommendation process.
Thus, we propose a privacy aware recommendation framework where users can freely choose which data to disclose with the recommender system.
As illustrated in \cref{fig:framework}, the critical difference between our framework and traditional recommendation is that the platform can only use the sub-data disclosed by the users. 
For example, the user on the left in \cref{fig:privacy_rec} can choose to hide his sensitive demographic attributes (e.g., age, gender, and education) and only discloses the last behavior to the service provider.

To enjoy the benefits of personalized services, users need to disclose their data to the recommender system to better model them.
Intuitively, more data the recommender system gets, better results the users can get.
However, disclosing data to the platform will increase users' privacy concerns, e.g., data abusing~\cite{Mayer:sp12:Third} and privacy leakage~\cite{zhang2021membership}.
Thus, under the privacy aware setting, users need to make information disclosure decisions based on the trade-off between anticipated privacy risks and potential utilities.
This idea can date back to \textit{Privacy Calculus Theory}~\cite{Laufer:si77:Privacy,Culnan:os99:Information}.

Before going into details, we first define the entire personal data $\di{}$ of user $i \in \mathcal{V}$'s as:
\begin{equation}
\di{} = \{{\scriptstyle \mathcal{D}_{i,a}, \mathcal{D}_{i,b}} \}= \{\{a_{i1},\dots, a_{iK}\}, \{b_{i1},\dots, b_{it_i}\}\},
\label{eq:di}
\end{equation}
where ${\scriptstyle \mathcal{D}_{i,a}}{=}\{a_{i1},\dots, a_{iK}\}$ denotes user $i$'s all profile attributes, $a_{ik}$ denotes the $k$-th profile attribute for user $i$, and $K$ is the number of profile attributes. ${\scriptstyle \mathcal{D}_{i,b}}{=}\{b_{i1},\dots, b_{it_i}\}$ denotes user $i$'s behaviors, $b_{ij}$ is the $j$-th behavior of user $i$, and $t_i$ is the last behavior timestamp. \czq{$\mathcal{V}$ is the set that includes all users in the platform.} %

A rational user is only willing to disclose data when she feels that she gains more from the platform than she loses in data disclosure.
Formally speaking, supposing user $i$ with whole data $\di{}$ currently discloses data $\si{} \subset \di{}$, now she tries to get a better recommendation results via disclosing more data $\si{'} \subset \di{}$ where $|\si{'}| > |\si{}|$, only if
\begin{equation}
    \texttt{U}_i(\si{'}) - \texttt{U}_i(\si{}) > \lambda_i\bigl(\texttt{C}_i( \si{'}) -\texttt{C}_i( \si{})\bigr),
\label{eq:base}
\end{equation}
where $\texttt{U}_i(x)$ denotes the utility that user $i$ can get from the platform with disclosed data $x$, function $\texttt{C}_i(x)$ measures the privacy cost paid by the user $i$ when she discloses the data $x$ to the platform, and $\lambda_i$ is the sensitive weight measuring how much user $i$ cares about her privacy.
Apparently, compared to privacy insensitive users (i.e., small $\lambda_i$), the platform needs to provide more performance improvements to attract privacy sensitive users (i.e., large $\lambda_i$) to disclose their data.
More details can be found in \cref{sec:user_type}.

\subsubsection{\textbf{User Objective.}}
Unlike traditional task settings where users can only passively accept recommendation results (i.e., without tools to optimize their objectives), in our framework, a rational user $i$ tends to maximize her utility $\texttt{U}_i(\si{})$ while minimize the privacy risk $\texttt{C}_i(\si{})$ by control the disclosed data $\si{}$.
The objective function for a specific user $i$ can be formalized as the following:
\begin{equation}
    \texttt{R}_i(\si{}) = -\lambda_i \texttt{C}_i(\si{}) + \texttt{U}_i( \si{}).
\label{eq:framework}
\end{equation}

The linear combination for user objective function follows the initial idea from \textit{Privacy Calculus Theory}~\cite{Laufer:si77:Privacy,Culnan:os99:Information} where both the recommendation performances from the platform and potential privacy cost are considered.
This formulation is also compatible with privacy related research in economics~\cite{farrell2012can,jin2017protecting,lin2019valuing}. %
They studied the micro-foundation on a user's intrinsic and instrumental preferences from disclosing personal information.
In our formulation, user's privacy cost $\texttt{C}_i(\si{})$ corresponds to intrinsic value for personal data (i.e., protecting the data from being obtained by others), while recommendation utility $\texttt{U}_i(\si{})$ corresponds to the instrumental value for personal data.

\subsubsection{\textbf{Platform Objective.}}
\label{sec:platform_obj}
In the proposed framework, the goal of a platform is still to maximize its revenue (e.g., purchases, clicks, or watching time) by improving the users' recommendation utility (e.g., providing more accurate results).
Thus, we define its objective as the summation of all users' recommendation utilities in \cref{eq:framework}\czq{, where $\mathcal{V}$ denotes all users in the platform}:
\begin{equation}
    \texttt{R}_{\text{p}} = \sum_{i \in {\scriptstyle \mathcal{V}}} \texttt{U}_i(\si{}).
\label{eq:platform}
\end{equation}
Considering the utility also depends on the recommendation model, the utility function $\texttt{U}_i(x)$ can be further defined as:
\begin{equation}
    \texttt{U}_i(\si{}) = \texttt{U}(\si{}) = \texttt{U}'(\si{},\, \texttt{M}_{{\scriptscriptstyle \mathcal{S}}}),
    \label{eq:updated_rec}
\end{equation}
where $\texttt{M}_{{\scriptscriptstyle \mathcal{S}}}: \si{} \rightarrow l_i$ ($l_i$ is recommendation results) is a recommendation model trained using all users' disclosed data ${\scriptstyle \mathcal{S}}=\{ {\scriptstyle \mathcal{S}_1},\dots, {\scriptstyle \mathcal{S}_{|\mathcal{V}|}}\}$ and $\texttt{U}'$ represents detailed recommendation utility function.
Here, without loss of generality, we assume that all users share the same utility function.
We will explore the personalized utility function in the future work.

\subsubsection{\textbf{Recommendation Utility Function}}
As shown in \cref{eq:framework} and \cref{eq:platform}, the recommendation utility $\texttt{U}$  occurs in the objective functions of both end users and the platform.
Here, we use the users' satisfaction with the results produced by the recommendation model to measure its utility.
It is worth noting that user satisfaction is still an open problem in \czq{recommender systems}.
\czq{Here, we simply quantify it by the user’s interactions} with the recommendation results, e.g., clicks, watches, and reads.
Based on such feedbacks, we can calculate different quantitative metrics as the utility in our framework, e.g., hit ratio and normalized discounted cumulative gain (NDCG)~\cite{ndcg}.
In this paper, we choose NDCG as the utility function $\texttt{U}$ for all users because of its widespread use \cite{NCF,kang2018self,Sun:cikm19:BERT4Rec}.

In traditional recommendation task, the platform can optimize this objective by only optimizing the model $\texttt{M}_{{\scriptscriptstyle \mathcal{S}}}$ since $\si{}= \di{}$ is a fixed, i.e., all users disclose their whole data.
However, this premise is broken in our proposed framework, where the user's disclosed data $\si{}$ is varying.
Thus, in our new framework, platforms also seek to attract users to share more data in other ways besides optimizing models, such as platform mechanism design.

\subsection{Platform Mechanism}
\label{sec:platform}

As mentioned before, the disclosed data $\si{}$ lives at the heart of the framework.
Ideally, user $i$ can freely choose any data $\si{}$ to disclose with the platform, e.g., choosing any profile attribute $a$ or behavior data $b$ as shown in \cref{fig:privacy_rec}.
However, in practice, such a degree of freedom is difficult to achieve for two reasons.
On the one hand, from the perspective of human-computer interaction, too fine granularity of disclosure choice (e.g., single behavior) can adversely hurt user experience~\cite{Zhang:hcs19:Proactive}.
On the other hand, although the privacy regulations ensure users the right to determine the use of their data, they do not stipulate how the service providers implement this function.

In practice, the platform usually designs some data disclosure mechanisms to provide the end users with several convenient options.
Here, we formulate the platform mechanism $\mathrm{G}=<\delta, \Pi>$ using two components, data split rule $\delta$ and disclosure choice spaces $\Pi$.
The data split rule $\delta$ is regarded as a function that reorganizes the original user data $\di{}$ using different granularity, and $\Pi$ denotes the space of all possible choices the platform provides to the user.
We illustrate a toy example in \cref{fig:mechanism}.

\subsubsection{\textbf{Data Split Rule}}
Since user data usually consists of two different data types (as in \cref{eq:di}), we defined $\delta$ as:
\begin{equation*}
    \delta (x) = \{\delta_a({\scriptstyle \mathcal{D}_{i,a}}), \delta_b({\scriptstyle \mathcal{D}_{i,b}}) \},
\end{equation*}
where $\delta_a$ and $\delta_b$ have similar forms that split the original data into several pieces according to the corresponding granularity and rules:
\begin{equation*}
    \{x_1, x_2,\dots,x_n\} \xrightarrow[\delta_a]{\delta_b}  \{x_1', x_2',\dots,x_m'\}, \,\, %
\end{equation*}
where $m \leq n$ and $x_j'$ is the candidate units for data disclosure. %
According to the segmentation rules, $x_j'$ can be several consecutive data points like $\{x_1, x_2, x_3\}$ or discontinuous random data like $\{x_5, x_{22}\}$.

$\delta_a$ aims to reorganize the user’s profile attributes.
The common approach is to keep original granularity (i.e., user can freely disclose any subset of attributes) or take all attributes as a whole (i.e., disclose all attributes or not).
Formally, it can be instantiated as:
\begin{equation*}
    \begin{aligned}
     \delta_a ({\scriptstyle \mathcal{D}_{i,a}}) &= \{a_{i1},\dots, a_{iK}\}\\
    \text{or}\quad  \delta_a ({\scriptstyle \mathcal{D}_{i,a}}) &= \{\{a_{i1},\dots, a_{iK}\}\}.
    \end{aligned}
\end{equation*}

Similarly, $\delta_b$ aims to transfer a user's original behavior data (e.g., thousands of clicks or more views) to few data disclosure options.
For example, ``percentage split'' with 10\% granularity divides a user's behavior sequence into 10 equal length subsequences, while ``daily split'' divides the user's behaviors by day.
Take ``percentage split'' with 10\% granularity as an example, it can be instantiated as: 
\begin{equation*}
    \delta_b ({\scriptstyle \mathcal{D}_{i,b}}) = \{{\scriptstyle \mathcal{S}_{i,b1}}, {\scriptstyle \mathcal{S}_{i,b2}}, \dots, {\scriptstyle \mathcal{S}_{i,b10}} \},
\end{equation*}
where ${\scriptstyle \mathcal{S}_{i,bj}} = \{b_{i,\lfloor 0.1 t_i*(j-1)\rfloor+1}, \dots, b_{i, \lfloor 0.1 t_i*j\rfloor}\}$ is the $j$-th candidate option of behavior data for user to disclosed.

\subsubsection{\textbf{Data Disclosure Choice Space $\Pi$}}

Assuming the platform has transferred user $i$'s original data $\di{}$ to $\delta(\di{}) = \{{\scriptstyle \mathcal{S}_{i,a1}},\dots,{\scriptstyle \mathcal{S}_{i,an}},$ $ {\scriptstyle \mathcal{S}_{i,b1}}, \dots,{\scriptstyle \mathcal{S}_{i,bm}}\}$, the platform can define data disclosure choice space $\Pi$ on these $m+n$ candidates as:
\begin{equation*}
    \begin{aligned}
    \Pi &= \{\Pi_1,\Pi_2, \dots, \Pi_N\}, \\
    \Pi_j & \sim [o_1, \cdots, o_k, \cdots, o_{n+m}], \quad o_k \in \{0,1\},
    \end{aligned}
\end{equation*}
where $o_k = 1$ denotes disclosing the $k$-th data in $\delta(\di{})$, while $o_k = 0$ means not; $\Pi_j$ is $j-$th data disclosure option that users can take; $N$ is the number of possible choices the platform provides to users.
For example, a full 0 vector $\Pi_j = [0, 0, \dots, 0]$ denotes that users can choose it to do not disclose any data.
More detailed instantiations can be found in \cref{sec:plat_mech}.

\tikzset{
  FARROW/.style={arrows={-{Latex[length=1.25mm, width=1.mm]}}},
  DFARROW/.style={arrows={{Latex[length=1.25mm, width=1.mm]}-{Latex[length=1.25mm, width=1.mm]}}},
  behavior/.style = {circle, fill=monte_carlo, minimum width=1.2em, align=center, inner sep=0, outer sep=0, font=\tiny},
  feature/.style = {circle, fill=salmon, minimum width=1.2em, align=center, inner sep=0, outer sep=0, font=\tiny},
  encoder/.style = {rectangle, fill=Madang!82, minimum width=6em, minimum height=3em, align=center, rounded corners=3},
  emb_layer/.style = {rectangle, fill=languid_lavender!72, minimum width=11em, minimum height=2em, align=center, rounded corners=3},
  project/.style = {rectangle, fill=hint_green, minimum width=7em, minimum height=2.4em, align=center, rounded corners=2},
  ds/.style={
       rectangle split,
       rectangle split part align=base,
       rectangle split horizontal=true,
       rectangle split draw splits=true,
       rectangle split parts=5,
       rectangle split part fill={athens_gray!80, athens_gray!80, athens_gray!80, athens_gray!80, athens_gray!80},
       draw=black, %
       very thin,
       minimum height=1.2em,
       minimum width=2em,
       text width=0.4em,
       inner sep=.5pt,
       text centered,
       font=\tiny,
       text=gray,
       },
}

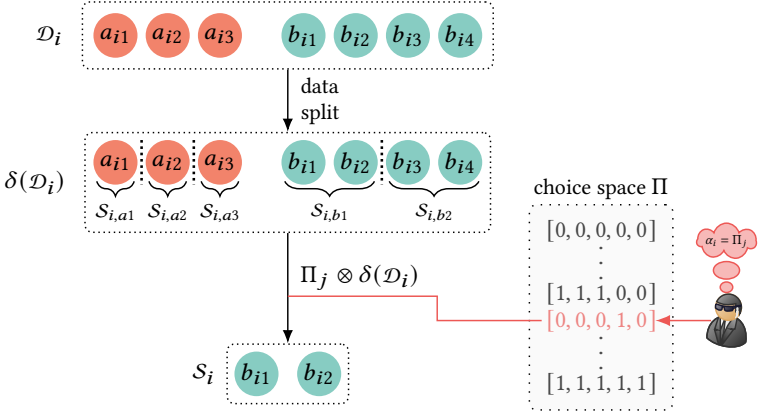
\begin{figure}
\centering
\resizebox{0.75\linewidth}{!}{
    \begin{tikzpicture}
        
    \node[] (mo)  at (0, 0) {};
    
    \node [feature, below of=mo, node distance=0cm, xshift=-0.9cm] (f1) {$a_{i1}$};
    \node [feature, right of=f1, node distance=0.5cm] (f2) {$a_{i2}$};
    \node [feature, right of=f2, node distance=0.5cm, ] (f3) {$a_{i3}$};
    \node [behavior, right of=f3, node distance=0.8cm, ] (b1) {$b_{i1}$};
    \node [behavior, right of=b1, node distance=0.5cm] (b2) {$b_{i2}$};
    \node [behavior, right of=b2, node distance=0.5cm] (b3) {$b_{i3}$};
    \node [behavior, right of=b3, node distance=0.5cm] (b4) {$b_{i4}$};
    
    \foreach \x in {1,2,3}
    { 
     \node [feature, below of=f\x, node distance=1.2cm ] (nf\x) {$a_{i\x}$};
     \node [behavior, below of=b\x, node distance=1.2cm ] (nb\x) {$b_{i\x}$};
    }
    \node [behavior, below of=b4, node distance=1.2cm ] (nb4) {$b_{i4}$};
    
    \foreach \x in {1,2}
    {
    \draw[densely dotted, line width= 0.7pt] ([xshift=2.5mm] nf\x.north) -> ([xshift=2.5mm]  nf\x.south);
    }
    \draw[densely dotted, line width= 0.7pt] ([xshift=2.5mm] nb2.north) -> ([xshift=2.5mm]  nb2.south);
    
    \draw [decorate, decoration={brace, amplitude=4pt, mirror}] ([xshift=-0.3mm, yshift=-0.3mm] nb1.south west) -- ([xshift=0.3mm, yshift=-0.3mm] nb2.south east) node[midway,yshift=-3mm, font=\tiny] (sb1) {$\scaleto{\mathcal{S}_{i,b1}}{5pt}$};
    \draw [decorate, decoration={brace, amplitude=4pt, mirror}] ([xshift=-0.3mm, yshift=-0.3mm] nb3.south west) -- ([xshift=0.3mm, yshift=-0.3mm] nb4.south east) node[midway,yshift=-3mm, font=\tiny] (sb2) {$\scaleto{\mathcal{S}_{i,b2}}{5pt}$};
    
    \foreach \x in {1,2,3}
    {
    \draw [decorate, decoration={brace, amplitude=4pt, mirror}] ([xshift=-0.3mm, yshift=-0.3mm] nf\x.south west) -- ([xshift=0.3mm, yshift=-0.3mm] nf\x.south east) node[midway,yshift=-3mm, font=\tiny] (sa1) {$\scaleto{\mathcal{S}_{i,a\x}}{5pt}$};
    }

    \node[densely dotted, draw=black, fit={(f1) (b4) }, inner sep=3, rounded corners=2] (d_box) {};
    \node[left = 0.5mm of d_box, font=\tiny] (di) {$\di{}$};

    \node[densely dotted, draw=black, fit={(nf1) ([xshift=-9mm, yshift=1mm] sa1.south west) (nb4) ([xshift=1.6mm, yshift=1mm] sb2.south east) }, inner sep=2, rounded corners=2] (nd_box) {};
    \node[left = 0.5mm of nd_box, font=\tiny] (di) {$\delta (\di{})$};
    
    \draw[FARROW] (d_box)-> (nd_box) node [pos=0.5, right, black, font=\tiny, align=left] {data\\split}; 
    
    \node[below = -2.5mm of nd_box, font=\tiny, xshift=3cm] (pi1) {$[0,0,0,0,0]$};
    \node[below of = pi1, font=\tiny, node distance =2mm] (pi2) {$\vdots$};
    \node[below of = pi2,  node distance =4mm, font=\tiny] (pi3) {$[1,1,1,0,0]$};
    \node[below of = pi3,  node distance =2.5mm, font=\tiny, text=flamingo] (pi4) {$[0,0,0,1,0]$};
    \node[below of = pi4, font=\tiny, node distance =2mm] (pi5) {$\vdots$};
    \node[below of = pi5,  node distance =4mm, font=\tiny] (pi6) {$[1,1,1,1,1]$};
    
    \node[dotted, draw=black, fit={(pi1) (pi6)}, inner sep=0.5, rounded corners=2, fill opacity=0.3, fill=gallery] (pi) {};
    \node[above = -0.8mm of pi, font=\tiny, xshift=0mm] (pi_l) {choice space $\Pi$};
    
    \node[maninblack, mirrored, minimum size=.2mm, right of=pi4, node distance=12mm] (ui) {};
    \draw[-latex, thin, flamingo] ([xshift=.4mm] ui.west)-> ([xshift=-1.5mm] pi4.east);
    \node [cloud, cloud puffs=7.2, cloud ignores aspect, draw=flamingo, fill=flamingo!42, minimum width=1mm, align=center, inner sep=0pt, above of=ui, node distance=0.75cm] (pii) {$\scaleto{\alpha_i=\Pi_j}{3.5pt}$};
    
    \node [ellipse, scale=0.5, draw=flamingo, fill=flamingo!42, minimum height=1.5mm, minimum width=5mm, above of=ui, node distance=0.95cm] () {};
    \node [ellipse, scale=0.3, draw=flamingo, fill=flamingo!42, minimum height=2mm, minimum width=4mm, above of=ui, node distance=1.05cm] {};

    \node [behavior, below of=nb4, node distance=2cm, xshift=-1.95cm] (sb1) {$b_{i1}$};
    \node [behavior, right of=sb1, node distance=0.6cm ] (sb2) {$b_{i2}$};
    
    \node[densely dotted, draw=black, fit={(sb2) (sb1)}, inner sep=2, rounded corners=2] (si) {};
    
    \draw[FARROW] (nd_box)-> (si) node [pos=0.4, right, black, font=\tiny, align=left] (pi_a) {$ \Pi_j \otimes \mathrm{\delta}(\di{}) $};
    
     \draw[thin, draw=flamingo] ([xshift=1mm] pi4.west) -- ++(-1, 0) -- ++(0, 0.23)  -- ++(-1.43, 0);
     
    \node[left of = si, font=\tiny, node distance=0.8cm] (si_l) {$\si{}$};

    \end{tikzpicture} }
    \caption{An illustrative example for platform mechanism. The platform firstly split the user's data $\di{}$ (three profile attributes and four behaviors) using the rule $\delta$ (keeping independent for attributes; percentage split with 50\% granularity for behaviors). Then it provides choices $\Pi$ for the user to choose to produce the final disclosed data $\si{}$.}
    \label{fig:mechanism}
\end{figure}

\subsubsection{\textbf{Platform Mechanism Design}}

With \czq{the} mechanism $\mathrm{G}=<\delta,\Pi>$, we can formally define the disclosed data $\si{}$ from user $i$.
Assuming user $i$'s original data $\di{}$ has been spited into candidates $\delta(\di{}) = \{{\scriptstyle \mathcal{S}_{i,1}}, \dots,{\scriptstyle \mathcal{S}_{i,m}}\}$\footnote{Here, we simplify the subscripts for easy description}. 
Then, $\si{}$ can be defined as the union of candidates in $\delta(\di{})$ selected  by a specific choice $ \alpha_i = \Pi_j \in \Pi$:
\begin{equation}
    \si{} = \alpha_i \otimes \mathrm{\delta}(\di{}) = \Bigl\{\scaleto{\bigcup_{\begin{subarray}{l}\,\,\, o_k=1\\\,\,\,  o_k \in \Pi_j\end{subarray}}}{20pt} {\scriptstyle \mathcal{S}_{i,k}} \Bigr\},
\label{eq:S_i}
\end{equation}
where $\delta$ is the platform data split rule and %
action $\alpha_i$ is sampled from user $i$'s privacy disclosure policy $\pi_i$, which decides the data to be disclosed. 
The operator $\otimes$ denotes the aggregation of the selected spilt data based on his choice $\alpha_i$.
\cref{fig:mechanism} illustrates a tiny example of the data disclosure process (i.e., generation process of $\si{}$) of a user with three profile attributes and four behaviors.

With formal definition of $\si{}$, we can  re-write the platform utility $\texttt{R}_{\text{p}}$ in \cref{eq:platform} using \czq{the} platform mechanism $\mathrm{G}$ and model $\texttt{M}_{{\scriptscriptstyle \mathcal{S}}}$ as below,
\begin{align}
     \texttt{R}_{\text{p}} |_{\mathrm{G=<\delta,\Pi>}} {=}  \sum_{i \in \mathcal{V}} \texttt{U}_i'(\si{}, \texttt{M}_{{\scriptscriptstyle \mathcal{S}}}) {=}   \sum_{i \in \mathcal{V}} \texttt{U}_i'(\alpha_i \otimes \mathrm{\delta}(\di{}) , \texttt{M}_{{\scriptscriptstyle \mathcal{S}}}).%
    \label{eq:platform_mechanism_optimization}
\end{align} 

One may figure out some possible optimal solutions towards the platform's best mechanisms.
However, the optimal platform mechanism design is another complex topic, usually considered from the view of game theory, and is out of scopes of this work.
Here, we take the first step, studying the data disclosure decision of users and platform revenues under several common mechanisms.

\subsection{Relationship with Privacy Preservation}

In this subsection, we will discuss the relationship between our work and privacy preservation, and clarify the position of our paper.

First, as claimed in the introduction, our work does not aim to propose a new method to directly protect the users' privacy data from privacy attacks.
The main motivation of this paper is to study the effectiveness of existing privacy mechanisms deployed in the platforms and further explore how different privacy mechanisms affect users' privacy decisions.
For this purpose, our proposed framework is model-agnostic.
The model $\texttt{M}_{\scriptscriptstyle \mathcal{S}}$ used here can be an ordinary recommendation model (e.g., NCF ~\cite{NCF} or GRU4Rec~\cite{Hidasi:ICLR2016:gru4rec}) or privacy-preserving recommendation models~\cite{Chen:TIST2020:Practical}.
For deploying a privacy-preserving model in our framework, we only need to modify the privacy cost function ($\texttt{C}_i$), taking into account the influence of protecting privacy, i.e., less privacy cost than a normal model when disclosing the same amount of data.
In this paper, for the convenience of analysis and considering that privacy-preserving technologies are not widely used in real-world applications, we only studied the proposed framework in ordinary recommendation models.
We leave the exploration of the impact of privacy-preserving technologies on users' privacy cost functions for future work. 

Second, our proposed framework provides users with the power to proactively trade off their privacy cost and recommendation utility.
From the final result point of view, users can optimize their data disclosure decision to discard those data that are not helpful for their recommendation results.
In this way, the proposed framework gives users the tool to achieve data minimization~\cite{Mireshghallah:WWW20:Not,biega2020operationalizing} by themselves, rather than waiting for the platform to implement the data minimization algorithms.
From this perspective, our proposed framework can be seen as  implicitly protecting user privacy.
Even if a privacy attack occurs, only part of users' disclosed data will be leaked.
Besides, giving users control over the recommendation process has also been found be effective in reducing their privacy concerns~\cite{Zhang2014-oa,Chen:CHI18:This}.

\section{Simulation}

\subsection{motivation} %
The most efficient way to figure out the answers to the questions we posed in the introduction is to deploy the proposed framework on a real-world platform and analyze how users adopt different and complex privacy policies to optimize their rewards.
However, direct deployment of these strategies and investments is currently impractical due to the following reasons.

Firstly, the most important reason is that such an online experiment may lead to the decline of the recommendation performances as well as the user experience, which harms the platform's revenue.
In the real world, nearly all the companies determine their platform mechanism driven by interest, and the revenues of the platforms are highly correlated with the recommendation performances. 
Therefore, it's nearly impossible to persuade any platform to directly deploy proposed strategies and mechanism online without other benefits.

Secondly, the experiments are \czq{built} upon several simplifications, mentioned in \cref{assumptions}, which poses challenges towards recommendation model training process.
For example, we assume when a user \czq{adjusts} his data disclosure policy, the recommendation system will forget his un-disclosed data. 
To facilitate such challenges, model unlearning or other privacy-preserving technologies are imposed.
However, in real-world applications, very few the e-commercial platforms have deployed these privacy-preserving technologies during the deep recommendation model training and evaluation processing.
As a result, we may still fail to guarantee the assumptions and simulation methodology becomes a substitution.

In summary, inspired by the success of simulation study on dynamic interactive problems in real-world applications~\cite{Ie:arxiv19:RecSim,krauth2020offline,lucherini2021t,yao2021measuring},  we employ the simulation to study the effects of the proposed framework and the possible game between users and the platform.

\subsection{Simplified Assumptions}
\label{assumptions}

To simplify the simulation process for easier analysis, we make some necessary assumptions to simplify the problem.

\begin{assumption}[Static Assumption] User $i$ optimizes her/his policy on the fixed data $\di{}$ which is not affected by user policy $\pi_i$.
\label{assumption:static}
\end{assumption}

Here static means the user data $\di{}$ is fixed during the simulation, but the disclosed data $\si{}$ produced by different user policies is dynamic. 
It is also the most common setting for recommendation task in research papers~\cite{Rendle:www10:Factorizing,Hidasi:ICLR2016:gru4rec,NCF,kang2018self,Sun:cikm19:BERT4Rec}.
In the simulation, we train the recommendation system $\texttt{M}_{{\scriptscriptstyle \mathcal{S}}}$ on the collected dynamic data $\mathcal{S}$ and validate the recommendation efficiency on a fixed test set. 
In real-world applications, the data $\di{}$, which contains the behavior data from the interaction with the recommender $\texttt{M}_{{\scriptscriptstyle \mathcal{S}}}$, is also dynamically changing with the user's policy $\pi_i$.
It is beyond the scope of this paper and we leave it as the future work.

\begin{assumption}[Immediate Assumption] The recommendation model $\texttt{M}_{{\scriptscriptstyle \mathcal{S}}}$ can only use the data $\si{}$ currently disclosed by each user $i$.
\label{assumption:forget}
\end{assumption}
The motivation of this assumption is that an untrusted platform can leverage user $i$' all data $\di{}$ if it can use the data disclosed in previous actions.
Without this constraint, the privacy right discussed in this paper is meaningless.
To achieve this, the platform can retrain the model from scratch with new data $\si{'}$ or quick unlearn the data in $\si{}$ then finetune with data $\si{'}$~\cite{cao2015towards,bourtoule2021machine,chen2022recommendation}.

However, the \cref{assumption:forget} also raises a new challenge that the asynchronous changes of user policy bring intractable computation costs for the platform since each time the user changes the disclosed data, the platform needs to update the model.
Here, we make an assumption for simplifying the simulation, assuming all users realize that the platform will cyclically (e.g., once a day) collect their privacy decisions and update recommender systems.
\begin{assumption}
[%
Cyclical Assumption]
Platform cyclically collects user privacy choices, and then the platform updates the model using all newly disclosed data. 
\label{assumption:synchronization}
\end{assumption}

In summary, for easy analysis in simulations, we introduce these assumptions to ignore the time and dynamic effects in this feedback system, just like the traditional recommendation task formulation.

\subsection{Platform Mechanism Simulation}
\label{sec:plat_mech}
In order to validate the effect of the platform mechanism, we adopt several mechanisms during simulation. 
For easy comparison, we utilize one mechanism at each experiment. %

\subsubsection{\textbf{Data Split Rule}}

\czq{In our simulation, we do not split the profile attribution and the user can determine whether to share all of their attributes.}
For behavior data, we apply ``percentage split'' as $\delta_b$ with different split granularity $p$ (e.g., 1/3) to split the behavior sequence into $1/p$ parts. 
One obvious advantage of ``percentage split'' is that it can normalize the size of user action space and decrease the inconvenience of the interaction between the user and the platform.

\subsubsection{\textbf{Data Disclosure Strategy}}
\label{sec:data_disclose_choice}

As the platforms have certain flexibility to implement different data disclosure strategies, we discuss three representative disclosure strategies used in our study for behavior data in this subsection.
These strategies determine the data disclosure action space $\Pi$ the user can choose.
For profile attributes, we found that all users tend not to disclose them in the experiments since these features are negligible for improving recommendation utility in the presence of behavior data.
Similar result that user profile features contribute very marginal to the recommendation results in the case of strong user behavior modeling on public benchmark datasets has also been reported in other works~\cite{kang2018self,Sun:cikm19:BERT4Rec}.
Thus, in the following study, we mainly focus on modeling only  behavior data.

The ``\textit{separate}'' rule gives the users the control to freely disclose any split personal data.
For this rule, the size of user $i$'s the action space is exponentially expended on the size of the spilt data set $|\delta_b ({\scriptstyle \mathcal{D}_{i,b}})|$, denoted as $2^{|\delta_b ({\scriptstyle \mathcal{D}_{i,b}})|}$. 
However, too many choices might make it difficult for users to make better data disclosure decisions.

Another data disclosure strategy named ``\textit{oldest continuous}'' provides users the choices to disclose continuous behavior data from the beginning time, such as selecting ``the oldest 33\% data''.
In this strategy, to disclose newer behavior data ${\scriptstyle \mathcal{S}_{i,bj}}$, users must also disclose all behavior data before it.
Take an already split data $\delta_b ({\scriptstyle \mathcal{D}_{i,b}}) = \{\scriptstyle \mathcal{S}_{i,b1}, \scriptstyle \mathcal{S}_{i,b2}, \scriptstyle \mathcal{S}_{i,b3}\}$ as an example, the action space provided by oldest continuous strategy is $\Pi = \{[0,0,0], [1,0,0], [1,1,0], [1,1,1]\}$, and its corresponding disclosed data is $\{\varnothing ,
    \{\! {\scriptstyle \mathcal{S}_{i,b1}} \!\},
    \{\! {\scriptstyle \mathcal{S}_{i,b1}} , {\scriptstyle \mathcal{S}_{i,b2}}\! \},$
    $\{ {\scriptstyle \mathcal{S}_{i,b1}}, {\!\scriptstyle \mathcal{S}_{i,b2}},$ ${\scriptstyle \mathcal{S}_{i,b3}}\} \}$.
``\textit{Latest continuous}'' mechanism is similar to ``oldest continuous'', with the only difference in the opposite direction.
The size of these two mechanisms' action spaces is $|\delta_b ({\scriptstyle \mathcal{D}_{i,b}})|$.

\subsection{User Policy Simulation}
\label{sec:user}

In this subsection, we introduce the simulation of user policy in our proposed framework.
As defined in \cref{eq:S_i}, the disclosed data $\si{}$ is result of the platform mechanism $\mathrm{G}$ and user's disclosure policy $\pi_i$. 
Meanwhile, in \cref{eq:updated_rec}, the recommendation utility $\texttt{U}_i( \si{})=\texttt{U}'(\si{},\, \texttt{M}_{{\scriptscriptstyle \mathcal{S}}})$ is also determined by the recommendation model $\texttt{M}_{{\scriptscriptstyle \mathcal{S}}}$, which is \czq{built} upon the all users' disclosed data $\mathcal{S}$. 
The reward of user $i$ may be varied even when $i$ keeps the disclosed data $\si{}$ unchanged since other users might change their disclosed data and the recommender system is changed.
Thus, the expectation rewards are considered rather than immediate value defined in \cref{eq:framework} and we assume all the users are rational and seek for the optimal privacy disclosure action  $\alpha_i^*$ to the optimal expected reward $E[ \texttt{R}_i | \alpha_i ] $ as his policy, i.e., %
\begin{equation}
    \begin{aligned}
    \alpha_i^{*} &= \argmax_{\alpha_i \in \Pi} E[ \texttt{R}_i | \alpha_i ]=  \argmax_{\si{\in [ \Pi \otimes \mathrm{\delta}(\di{}) ] }} E[ \texttt{R}_i(\si{)} ] \\
& =\argmax_{\alpha_i \in \Pi } E\Bigl[ -\lambda_i \texttt{C}_i\bigl( \alpha_i \otimes \mathrm{\delta}(\di{}) \bigr) + \texttt{U}_i\bigl( \alpha_i \otimes \mathrm{\delta}(\di{}) )\bigr) \Bigr].
    \end{aligned}
\label{eq:opt_pi}
\end{equation}

As mentioned before, recommendation utility $\texttt{U}_i$ has been discussed in \cref{sec:platform_obj}.
To study this objective, we need to define the privacy cost function $\texttt{C}_i$ and sensitive weight $\lambda_i$.

\subsubsection{\textbf{Privacy Cost Function}}
\label{sec:privacy_cost}
We simulate every user with the same cost function $\texttt{C}$ and leave the diversity of user privacy sensitivity to the parameter $\lambda_i$. 
Following current experiment specifications in the economics literature~\cite{lin2019valuing,tang2019value}, we model the privacy cost function as a linear summation\footnote{See the Eq. 2 in \cite{lin2019valuing} and the dis-utility from disclosure in the econometric specification session in \cite{tang2019value}.} of disclosed personal data.

\czq{According to the comprehensive survey on privacy value definition \cite{MKT-053}, people will measure the value of their privacy into the intrinsic value of privacy and the instrumental value of privacy.}
\czq{
The intrinsic loss indicates the sake of protecting their intrinsic private data, which measures the valuation on the intrinsic properties such as the education or the income levels. }
\czq{In this work, we denote the intrinsic loss towards the privacy cost on amount of the sharing user profile attributes.}
\czq{The instrumental value of privacy indicates how the transaction efficiency would be affected by sharing user data, especially the data generated in the applications. 
In this work we denote the privacy cost towards the percentage of shard user historical behavior data. 
Therefore, the privacy cost function is described below,}
\begin{equation}
    \texttt{C}_i(\si{})= \beta_i * | {\scriptstyle \mathcal{S}_{i,a}} | + \frac{| {\scriptstyle \mathcal{S}_{i,b}} |}{ | {\scriptstyle \mathcal{D}_{i,b}} |}
    \label{eq:cost_function0}
\end{equation}
\czq{where the first term indicates the intrinsic loss and the second term indicates the instrumental loss.} 
\czq{If user does not tend to disclose profile attribute, such privacy cost function can be simplified to the following format with the instrumental value alone.}
As mentioned in \cref{sec:plat_mech}, user tends not to disclose profile attributes $\scriptstyle \mathcal{D}_{i,a}$ due to no gains in our experiments, so we only consider behavior data here, i.e.,
\begin{equation}
    \texttt{C}_i(\si{})=\texttt{C}(\si{}) =  \frac{| {\scriptstyle \mathcal{S}_{i,b}} |}{ | {\scriptstyle \mathcal{D}_{i,b}} |}
    , %
    \label{eq:cost_function}
\end{equation}
where the $|x|$ is the number of elements in $x$.
Here, the percentage based measurement regards different amount of users' data equally. %

This reduced form specification is not unrealistic as it captures the substitution effect among personal data and incorporates the idea of constant marginal privacy cost. 
One might argue for a higher order functional to capture richer implications. 
However, there is little experimental evidence that the higher order form for privacy cost exists and how the functional form looks like.

\subsubsection{\textbf{Privacy Sensitive Weight}}
\label{sec:user_type}
For user $i$ who disclosed all her/his data (i.e., $\si{} = \di{}$), her/his privacy cost compared to not sharing any data (i.e., $\si{} = \varnothing $) is 
\begin{equation}
    \texttt{C}(\di{}) - \texttt{C}(\varnothing).
\label{eq:privacy_diff}
\end{equation}
Meanwhile, her/his anticipated recommendation utility compared to not sharing any data is:
\begin{equation}
    \texttt{U}(\di{}) - \texttt{U}(\varnothing).
\label{eq:utility_diff}
\end{equation}
We assume all users have accessed to the recommendation utility $\texttt{U}(\di{})=\texttt{U}'(\di{},\texttt{M}_{{\scriptscriptstyle \mathcal{D}}})$ computed on all the data $\di{}$ and the recommendation utility without their data $\texttt{U}(\varnothing)$ before they can take data disclosing actions, which can be regard as a prior knowledge, like the experiences before the platform adopted our framework.
With \cref{eq:privacy_diff} and \cref{eq:utility_diff}, we define the privacy sensitive weight $\lambda_i$ as: 
\begin{equation}
    \lambda_i = w_i  * \frac{\texttt{U}(\di{})  - \texttt{U}(\varnothing) } {  \texttt{C}(\di{}) - \texttt{C}(\varnothing) },
    \label{marginal_define}
\end{equation}
where $w_i$ indicates the diversity of user types towards privacy sensitivity.
The users with $w_i > 1$ is privacy sensitive users, as they will not be willing to disclose the corresponding data $\di{}$ if they only get $\texttt{U}(\di{})$ as before.
While users  with $w_i < 1$ are just the opposite. %
Therefore, the user's privacy sensitive weight is pre-computed, and the $\texttt{U}(\di{})$ can be regarded as the benchmark expectation of the platform.
The formulation of the privacy sensitive weight $\lambda_i$ also meets the idea from \cite{lin2019valuing}, where the heterogeneity from users' social demographic variety should also be explicitly characterized. %

\subsubsection{\textbf{Simulation Algorithm}}

As users behave rationally to find the optimal strategy with a trade-off of exploration and exploitation, it just meets the idea of the reinforcement learning algorithm. 
Therefore, we model each user as a unique agent and apply a multi-agent reinforcement learning method to simulate user possible policy adaptation. 
The recommender system is regarded as the environment to provide feedback, which is built upon the disclosed user data.
All agents' policies are optimized simultaneously by determining their actions, i.e., the disclosed data $\scriptstyle \mathcal{S}^t$ at simulation epoch $t$, which is used to train the recommendation model $\texttt{M}_{{\scriptscriptstyle \mathcal{S}}^{t}}$.
As mentioned before, users tend to find an optimal action over possible action space $\Pi$ to maximize his expected reward, which is determined by all agents in this dynamic MARL environment.

We assume each user (agent) realizes this situation that the immediate reward is the result of all agents, but no communication or observation among agents is permitted. 
Then, each agent is concerned about her/his own utility and regards the environment as a dynamic system that is partially correlated to herself/himself. 
Now, it is simplified to a Multi-Armed Bandit problem~\cite{katehakis1987multi}.

However, the challenge of the exploration and explication problem also exists in our simulation. 
To address it, we adopt a simple but efficient method, Epsilon Greedy~\cite{sutton2018reinforcement} algorithm, to simulate user's policy $\pi_i$ as following, %
\begin{equation}
     \alpha_i^{t+1} = \left\{
\begin{array}{l l }
\alpha_i \sim \texttt{P}^t_i,
& \text{with possibility } \epsilon \\
\argmax_{\alpha_i} Q_i^t(\alpha_i), & \text{with possibility }  1-\epsilon \\
\end{array} \right. 
    \label{epsilon_greedy}
\end{equation}
where $Q_i^t(\alpha)$ is the user $i$'s estimation value at simulation epoch $t$ on action $\alpha$, and $\texttt{P}^t_i$ denotes a random sample policy.
To conduct an efficient policy exploration, we sample a less explored action with a higher possibility as following,
\begin{equation}
    \texttt{P}^t_i(\alpha) = \frac{ 1/ (N^{t-1}_i(\alpha) +1) } { \sum_{x \in \Pi} 1/(N^{t-1}_i(x) +1) },
    \label{random_rule}
\end{equation}
where $N^{t-1}_i(\alpha)$ represents the total number of action $\alpha$ was taken by user $i$ from start to the last simulation epoch $t{-}1$.
In convenience, we adopt the approximated expected estimation results and 
update it with the residual between the estimation $Q_i^{t-1}(\alpha_i^{t-1})$ and immediate reward  $\texttt{R}_i^{t-1}$ when she/he takes action $\alpha_i^{t-1}$ as following.
\begin{equation*}
       Q_i^t(\alpha) {=} \left\{\!\!\!
\begin{array}{l l}
Q_i^{t-1}(\alpha), &  \text{if } \alpha_i^{t-1} {\neq} \alpha \\
Q_i^{t{-}1}(\alpha) {+} \frac{1}{N^t_i(\alpha)} \bigl(\texttt{R}_i^{t-1}(\! \alpha {\otimes} \mathrm{\delta}(\di{})  ) {-} Q_i^{t{-}1}(\alpha) \bigr), &  \text{if }  \alpha_i^{t{-}1} {=} \alpha \\
\end{array} \right. 
\end{equation*}
where $\texttt{R}_i^{t-1}$ is user $i$-th immediate objective at simulation epoch $t{-}1$, computed by \cref{eq:framework}. 
$Q_i^0(\alpha) $ is the user $i$'s initial expected reward if she/he takes action $\alpha$. 
which is initialized to $0$ as users have no prior about their behaviors on the new dynamic environment.

In our simulation, we set initial $\epsilon=0.5$ for all agents and decay a half during the MARL training processing. The detailed decay epoch is co-related to the size of possible action space $\Pi$.
Here, we define it as $\epsilon = 0.5^{ t /(3 * |\Pi |) }  $, 
where $t$ is the epoch during the reinforcement learning training processing. 

\subsection{\textbf{Discussion}}
To figure out how the platform mechanism affects users' behavior, we turn to the simulation built upon several simplified assumptions. 
One fundamental assumption is the hypothesis of rational man, where users will seek their optimal policies to maximize their objectives. 
However, in the real-world scenarios, human behaviors are also affected by psychological factors, which should also be modeled in future work.
One detailed example is that some users may feel exhausted digging out all the potential privacy choices with the provided platform mechanism.
In our simulation, we assume there remains no mental cost when a user adjusts his policy. 
However, in the reality, some users may refuse to change their policy frequently, especially in complex user interaction applications.
For such situation, a convenient user interface (UI) could be a potential solution to mitigate users' fatigues. 
\czq{
Another important factor is that users may adjust their trust level towards the platform during their exploration. One detailed example is that if the platform or even the recommender system \cite{zhang2022pipattack} is easy to be attacked or the platform will abuse their disclosed data to other applications, they may re-consider their privacy sensitivity. 
Though some works have discussed the utilization of trusted platform or the privacy-preserved recommendation model, the possible effects on user psychological factors might be tackled by a dynamic modeling on the user privacy sensitive weights, which is out of the scope of this work.
}
We simplify the influences of the psychological factors in this work and leave the exploration of psychological effects in mechanism designs and UI designs for future works.

\section{Experiments}

\subsection{Research Question}
\begin{itemize}[itemsep=5pt, topsep=12pt,]
    \item \textbf{RQ1}: How do different platform mechanisms affect the recommendation performance and the data disclosure decisions of users with different privacy sensitivity?
    \item \textbf{RQ2}: What is the role of recommendation model in this framework? Can a more accurate model attract users to disclose more data?
    \item \textbf{RQ3}: How does user population composition affects the user behavior in this framework?
\end{itemize}

\subsection{Experiment Setup}

\begin{table}
    \centering
    \caption{Statistical details of the evaluation datasets.}
    \label{tab:dataset}
    \begin{tabular}{lrrrrr}
\toprule
Dataset & \#User & \#Item  & \#Interaction & Density \\ \midrule
ML-100k &637&   1278   & \num{90554}   & 11.12\%   \\ 
Yelp &    8338 &  \num{35476}   &  \num{760635}  & 0.26\%    \\
\bottomrule
\end{tabular}
\end{table}

\subsubsection{Dataset}
 We conduct our experiments on two real-world representative datasets which vary in domains and sparsity:
\begin{itemize}
    \item \textbf{Movielens-100k}\footnote{\url{https://grouplens.org/datasets/movielens/100k/}} (ML-100k) \cite{Harper:TIIS16:MovieLens}: This is a popular benchmark dataset for evaluating recommendation algorithms. Here, we use the version that includes 100k user ratings.
    \item \textbf{Yelp}\footnote{\url{https://www.yelp.com/dataset}}. This is a popular and continuously updated dataset for business recommendation. The version we download for the experiments in this paper is Feb. 16 2021, i.e., all review records are written before Feb. 16 2021.
\end{itemize}

 Since we focus on recommendation based on implicit feedback, we follow the common practice to convert the numeric rating or a review into implicit feedback of 1 (i.e., indicating the user interacted with the item). %
 After that, we build the behavior sequence for each user by grouping and sorting their behaviors according to the timestamps.
 To properly simulate the information disclosure decision making process, we filter out the user with less than 40 interactions and items with less than 5 interactions.
 For efficiency reasons, we further subsample the users in Yelp, resulting in a dataset with 8338 users.
The statistics of the processed datasets are summarized in \cref{tab:dataset}.

\subsubsection{Simulation Setup}

For each user, we hold out the last item of the behavior data as the test data to compute recommendation utility \cite{NCF,kang2018self,Sun:cikm19:BERT4Rec}.
The rest of the behavior data is used for training simulation, treating the last interaction data in disclosed data as validation data and the remaining disclosed data for training data.

For recommendation utility evaluation, we adopt the widely used \textit{leave-one-out evaluation} \cite{NCF,kang2018self,Sun:cikm19:BERT4Rec} protocol with  NDCG@100 computed on the whole item set as the metric.
In particular, for Yelp, we compute the sampled metric with 1000 negative samples since the large item candidate set makes the results on Yelp are too small to simulate stably.
These sampled results are consistent with the scores on the whole candidate set \cite{krichene2020sampled}.

For privacy risk function, we use disclosed data percentage measurement according to \cref{sec:privacy_cost}, which is weighted by the sensitive weight $\lambda_i$ defined in \cref{marginal_define}.
To study the data disclosure decision making for users with different privacy sensitivity, we randomly divide users into three groups (each with $1/3$ users) with different privacy sensitive levels by adjusting the $w_i$, 
\begin{itemize} %
    \item $w_i = 0$: non-sensitive user who does not care privacy at all.
    \item $w_i = 1$: normal user who weights privacy risk and recommendation results in a relatively normal way.  
    \item $w_i = 10$: sensitive user who concerns more about privacy than recommendation utility.
\end{itemize}
To acquire the sensitive weight $\lambda_i$, the benchmark recommendation result $\texttt{U}(\di{})$ is computed based on GRU model with the whole dataset.
The assumption here is that the non-privacy aware framework that the user used before was based on GRU4Rec.
\czq{As this work focus on the effects on the user disclose choices, we also assume that there exists no privacy leak among data transmission period and users only access to their private recommendation results and the platform recommendation models are well protected with privacy guarantee. 
}
It is worth noting that we group users into three categories here just for the convenience of analyses and discussions in subsequent experiments.
In fact, according to \cref{marginal_define}, users in each category still have different privacy sensitivity $\lambda_i$.  

In the RL training, we model each user as an agent following the setup, and train 400 epochs with Epsilon Greedy algorithm. 
In each simulation epoch, the recommendation model is trained from scratch as discussed in \cref{assumption:forget}, i.e., the platform can only use the data that the user disclosed in the current simulation epoch.
The simulation epoch is enlarged to 3000 epochs in
``separate'' data disclosure strategy with $p=1/8$ due to the slow convergence.

\subsubsection{Recommendation Model}
To study the role of different models in users' data disclosure decision making, we conduct the simulations on different models, including two state of the art sequential recommendation models  and one CF model.
\begin{itemize}
    \item \textbf{GRU4Rec}~\cite{Hidasi:ICLR2016:gru4rec}: It uses GRU with ranking based loss to model user sequences for session based recommendation. 
    \item \textbf{BiSA} (\textbf{Bi}directional \textbf{S}elf-\textbf{A}ttention) \cite{kang2018self,Sun:cikm19:BERT4Rec}: It uses a self-attention architecture to capture users’ sequential behaviors and  achieves state-of-the-art performance on sequential recommendation. It usually can obtain better results than GRU4Rec with more powerful architecture.
    \item \textbf{NCF} \cite{NCF}: NCF models user–item interactions with a multi layer perceptron. It is included as a weaker model since it is not designed to capture the sequential information in user \czq{behaviors}.
\end{itemize}

We implement these models using \texttt{PyTorch}\footnote{The source code will be released after the review phase.}.
The \czq{hyper-parameters} are carefully tuned using a grid search to achieve optimal performances.
After tuning, the embedding size and hidden size is set to 128 for all the models, the dropout ratio is set to 0.2, the learning rate equals to 1e-3 for the models except BiSA with a learning rate 3e-4, and the number of negative samples is set to 16.
All models are trained with adam optimizer~\cite{kingma2015adam} with early stop.

\subsection{Study 0: Impact of User Profile Attribution}
\czq{Before conduct our experiment, we validate whether user feature perform a essential part of our latter experiments with Movielens-100k for example.
In Movielen, we include all provided 4 user profile attributes: sexual, age, job, and zip code.}
\czq{We firstly conduct experiments on the situation where all users submit their historical data with and without user profile with three mentioned models in the following tables}
\begin{table}
    \centering
    \caption{The averaged recommendation results (\%) on three different recommendation models when all users disclose history behaviors with or without profile attributes.
    }
    \label{tab:profile_rec_results}
    \begin{adjustbox}{max width=\linewidth}
    \begin{tabular}{@{}llrrr@{}}
\toprule
{Dataset} & {Model} & {with profile}&{without profile} \\ 

\midrule
\multirow{3}{*}{{ML-100k}} & {NCF} & 12.69 & 12.65   \\ 
& {GRU} &   19.89  &  19.87     \\
& {BiSA} &   19.90  & 20.06    \\

\bottomrule
\end{tabular}
\end{adjustbox}
\end{table}
\czq{From the \cref{tab:profile_rec_results} we observe that the user profile attributes introduce minor recommendation improvements on all three models if the recommendation systems have already absorbed enough user information from their historical data.}
\czq{In other words, when the platforms have already collected enough information for the recommender systems, users have minor incentives to disclose their user profiles attributes.}
\czq{we secondly conduct experiments to simulate the circumstances where all users are able to determine whether to disclose their history behavior data along with their profile attributes.}
\begin{table}
    \centering
    \caption{The percentage(\%) of the users who are willing to disclose behavior data or the profile data under different split granularity after convergence. 
    }
    \label{tab:profile_gru_results}
    \begin{adjustbox}{max width=\linewidth}
    \begin{tabular}{@{}llrrr@{}}
\toprule
{Data Composition} & {p} & {Disclose behavior percentage}&{Disclose profile percentage} \\ 

\midrule
\multirow{3}{*}{{With Profile}} & {1} & 42.39 & 6.00   \\ 
& {1/2} &   52.90  &  6.28     \\
& {1/4} &   60.91  & 6.28    \\
\midrule
\multirow{3}{*}{{Without Profile}} & {1} & 43.80 & -   \\ 
& {1/2} &   52.74  &  -    \\
& {1/4} &   56.67  & -    \\

\bottomrule
\end{tabular}
\end{adjustbox}
\end{table}
\czq{In the following experiments, the guarantee that the recommender system collect sufficient user information may not hold, where we conduct simulations with the ``continuous'' data spilt rule on GRU models under 3 different granularity. 
In simplify, we set the $\beta_i=1/|{\scriptstyle \mathcal{D}_{i,b}} |$ according the \cref{eq:cost_function0}.
As we import the user profile attribution, the action spaces of agents are enlarged, which impose the convergence challenges.
Therefore, we restrict the users whose disclosure decision have at least 0.01 improvements on their utility than not to disclose.
The converged results are displayed on the \cref{tab:profile_gru_results} where we record the users who are willing to disclose their historical behavior data or their profile attributes.
From \cref{tab:profile_gru_results}, we can conduct two conclusions: First, barely users are stick to disclose their profile attribution among different data split granularity;
Second, the user disclosure policies towards their historical data are similar no matter their profile attribution is introduced or not. The converged number of the users who are willing to disclose their historical are similar from \cref{tab:profile_gru_results}. Regarding the slight influences on the user profile attributes, the introduce of the user profile attributes will directly enlarge the user possible action space, which impose more challenge on the multi-agent reinforcement learning processing and may require much more training cost. }

\czq{As a results, we only conduct experiments on user historical data and assume all users are refuse to disclose their user profile attribution on the latter experiments. 
}

\subsection{Study 1: Impact of Platform Mechanism}

We firstly conduct experiments on platform mechanisms specifying various data split granularity and data disclosure strategies with a widely-used sequential recommendation model GRU4Rec. 
We begin by answering which mechanism is preferred by users with different privacy sensitivity (types, hereafter).

\begin{table*}
\centering
\caption{Results on different data sets with different platform mechanisms.
    All the results are averaged on the last 20 epochs.
    ``dis.\%'' denotes average percentage of disclosed data , ``NDCG'' means NDCG@100 (\%).}
\label{Rec_results}
\renewcommand{\arraystretch}{1.02}
\begin{adjustbox}{max width=\textwidth}
    \begin{tabular}{l r r r r r r r r r} \toprule 
        &   & \multicolumn{8}{c}{ML-100k} \\
         \cmidrule(lr){3-10} 
        strategy & \multicolumn{1}{c}{$p$} & \multicolumn{2}{c}{non-sensitive}  & \multicolumn{2}{c}{normal} & \multicolumn{2}{c}{sensitive} & \multicolumn{2}{c}{all} \\
        \cmidrule(lr){3-4}  \cmidrule(lr){5-6} \cmidrule(lr){7-8}  \cmidrule(lr){9-10} 
        & & NDCG & dis.\% & NDCG & dis.\%  & NDCG & dis.\%  & NDCG & dis.\%  \\ 
        \midrule
        all & \multicolumn{1}{c}{---}  &  21.74 & 100 & 19.45 & 100 & 17.46  & 100 &   19.45 &  100 %
        \\ 
        \midrule
       \multirow{5}{*}{\shortstack[l]{latest\\ continuous}} 
       &  1  &
      17.07 &  100 & 10.07 & 21.80  & 6.72 & \underline{11.75}  & 11.30 & 43.37  %
      \\
       &  1/2  &
      17.54 &  100 & 16.71 & \underline{25.11}  &  7.38 & 8.88 & 13.89 & \underline{43.57}  %
      \\
       & 1/4  &
     \textbf{18.19} &  100 &  18.54 & 18.39 & 6.99 & 5.75   & 14.58 & 40.19 %
     \\
        & 1/8  & 
       17.31 &  100 &  19.04 & 17.83 & 8.23 & 7.08 & 14.87 & 40.43   %
      \\
       & 1/16  & 
      17.65 &  100 &  \underline{19.26} & 16.22 & \underline{12.01} & 8.82 & \underline{16.31} &  40.45  %
     \\
        \midrule
       \multirow{5}{*}{\shortstack[l]{oldest\\ continuous}} 
       & 1  &
      17.07 &  100 & 10.07 & 21.80  & 6.72 & 11.75  & 11.30 & 43.37  %
      \\
       &  1/2  &
      17.40 &  100&  14.76 & 25.99 &  8.07 & 11.98  & 13.41 & 44.91  %
      \\
       &  1/4  &
       17.48 & 100 &  16.10 & \textbf{28.67} & 8.63 & 12.96 & 14.07 &  46.17 %
       \\
        &  1/8  & 
       17.44 & 100 &  18.08 & 22.63 & 10.35 & 15.41 & 15.28 & 44.78 %
      \\
       & 1/16  & 
       \underline{17.55} & 100 &  \underline{21.55} & 26.85 & \underline{12.88} & \textbf{17.03} & \underline{17.33} &  \textbf{46.89}  %
       \\
        \midrule
       \multirow{4}{*}{\shortstack[l]{separate}} 
       &  1  &
    17.07 &  100 & 10.07 & 21.80  & 6.72 & 11.75  & 11.30 & 43.37   %
      \\
       & 1/2  &
      17.91 & 100 &  19.35 & \underline{28.57} &  8.06 & 11.05  & 15.11 & 45.50 %
      \\
       &  1/4  &
       17.38 & 100 &  22.26 & 23.88  & 9.44 & 11.85 & 16.36 & 44.12 %
        \\
        & 1/8  & 
       \underline{18.08} & 100 &  \textbf{27.63} & 25.65  & \textbf{14.93} & \underline{17.00} & \textbf{20.21} & \underline{46.46} %
      \\
    
    \toprule 
        &   & \multicolumn{8}{c}{Yelp} \\
         \cmidrule(lr){3-10} 
    \multirow{5}{*}{\shortstack[l]{latest\\ continuous}} 
       &  1  &
      26.59 & 100 & 11.36 &32.58  & 3.13  & \underline{6.11} & 13.70 & 46.70  \\
       &  1/2  &
      \textbf{27.23} & 100 &  24.38 & \textbf{}{41.45} & 3.48 & 5.49 & 18.37 & \underline{49.43} \\
       & 1/4  &
      26.85 & 100 & 26.72 & 27.25 & 3.93  &  5.07 &  19.17 & 44.60 \\
        & 1/8  & 
      26.28 & 100  & \underline{27.00} & 19.49 & 7.44 & 5.49 &  20.24 & 42.18 \\
       & 1/16  & 
     26.09 &  100 & 25.82 & 17.22 & \textbf{15.62} & 6.08 &  \underline{22.33} &  41.10 \\
        \midrule
       \multirow{5}{*}{\shortstack[l]{oldest\\ continuous}} 
       & 1  &
      \underline{26.59} & 100 & 11.36 &32.58  & 3.13  & 6.11 & 13.70 & 46.70  \\
       &  1/2  &
      26.47 & 100 & 19.34  & \underline{37.27} & 3.15  & 5.84 & 16.32 &  \underline{48.16} \\
       &  1/4  &
      26.16 & 100 & 22.25 & 30.91  & 3.60 & 5.42  & 17.34 &  45.92 \\
        &  1/8  & 
      26.02 &100 & 23.76  & 26.31 &  5.49 & 5.64 & 18.42 & 44.49 \\
       & 1/16  & 
       25.99 & 100 & \underline{24.77} & 24.68 & \underline{11.06} & \underline{6.46} & \underline{20.61} & 44.21    \\
        \midrule
       \multirow{4}{*}{\shortstack[l]{separate}} 
       &  1  &
      26.59 & 100 & 11.36 &32.58  & 3.13  & 6.11 & 13.70 & 46.70  \\
       & 1/2  &
      \underline{27.00} & 100 & 26.86 & \textbf{42.91}  & 3.79 & 4.71 & 19.22 &  \textbf{49.65} \\
       &  1/4  &
       26.48 & 100 & \textbf{29.97} & 27.73  & 4.56  & 4.64 & 20.34 & 44.61  \\
        & 1/8  & 
      26.60 & 100 & 29.14 & 27.97  & \underline{13.97} &  \textbf{15.71} & \textbf{23.23} &  48.34 \\
        \bottomrule
    \end{tabular}
    \end{adjustbox}
    
\end{table*}

\subsubsection{Split Granularity $p$}
We first validate how the split granularity affects users on the three aforementioned data disclosure strategies.
The recommendation results and the data disclosure percentage on different user types are reported in \cref{Rec_results} where all results are averaged on the last 20 epochs after convergence.
From the results, it can be observed that:
\begin{enumerate} [itemsep=4pt]
    \item Comparing the NDCG performances among different settings, we can derive a negative answer for the question in the introduction considering ``all or nothing'' binary choice (i.e., $p{=}1$) performs worst on all disclosure strategies for all datasets.
Even looking at the detailed results for user groups with different privacy sensitivity, $p{=}1$ still performs very poorly, if not the worst.
\item Comparing the results within different user groups, a prominent and expected result is that users who care about privacy are only willing to disclose very little data, especially privacy sensitive users.
Besides, normal users can obtain comparable recommendation results (even better on ML-100k) to non-sensitive users with much less data.
On the one hand, this indicates that our proposed framework can effectively protect users' privacy.
On the other hand, proactively controlling the disclosed data also allows users to improve their recommendation results by themselves.
\item For platform, finer split granularity can usually bring better performances in all three mechanisms for all datasets.
Unexpectedly, these superior performances are not always obtained through more disclosed data.
For example, under ``latest continuous'' rule, the overall recommendation performances for $p{=}1/16$ (16.31\% on ML-100k) are much better than $p{=}1/2$ (13.89\% on ML-100k) with less training data (40.45\% vs. 43.57\% on ML-100k).
\end{enumerate}

To figure out the reason for this phenomenon, we analyzed the distribution of user' data disclosure.
We reported the percentage of users who disclosed data in \cref{tab:sharing_result}.
The results show that more users turn to disclose data since finer granularity allows users to disclose a small amount of data for certain recommendation utilities.
Conversely, more users suffer from poor recommendations as they refuse to disclose data under coarse-grained granularity.

\begin{table}
    \centering
    \caption{The percentage (\%) of users who disclosed data, which is averaged on the last 20 epochs.
    }
    \label{tab:sharing_result}
    \begin{adjustbox}{max width=\linewidth}
    \begin{tabular}{@{}llrrrrrr@{}}
\toprule
{Dataset} & {Strategy} & {\#p=1}&{\#p=1/2} & {\#p=1/4}  & {\#p=1/8} & {\#p=1/16} \\ 

\midrule
\multirow{3}{*}{{ML-100k}} & {latest} & 43.80 & 52.74 &   56.67   & 61.02   & 70.63   \\ 
& {oldest} &   43.80  &  50.39  &  51.81  & 59.65 & 64.65   \\
& {separate} &   43.80  & 56.54 &  66.27   &  79.48  & -    \\
\midrule
\multirow{3}{*}{{Yelp}} & {latest} & 45.90 & 62.03 &  66.34   & 70.11   & 79.64   \\ 
& {oldest} & 45.90 & 56.73  & 61.37   & 64.94  & 73.33    \\
& {separate} &45.90  & 63.68   &  68.26  & 80.82 &-    \\
\bottomrule
\end{tabular}
\end{adjustbox}
\end{table}

\subsubsection{Data Disclosure Strategy}
\pgfplotsset{
axis background/.style={fill=gallery},
grid=both,
  xtick pos=left,
  ytick pos=left,
  tick style={
    major grid style={style=white,line width=1pt},
    minor grid style=gallery,
    draw=none
    },
  minor tick num=1,
  ymajorgrids,
	major grid style={draw=white},
	y axis line style={opacity=0},
	tickwidth=0pt,
}

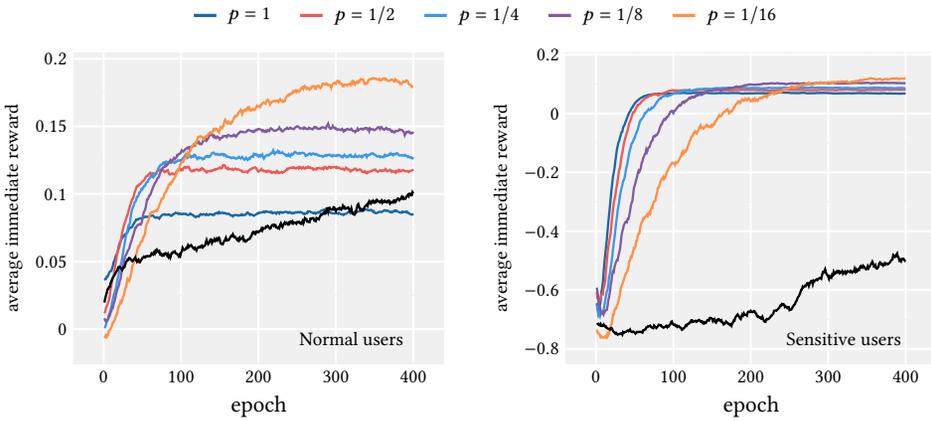
\begin{figure}
\centering
\begin{tikzpicture}[scale=0.72]
\begin{groupplot}[
	    group style={group size=2 by 1,
	        horizontal sep = 64pt,
	        },
	    xlabel= \Large epoch,
        ymajorgrids,
        major grid style={draw=white},
        y axis line style={opacity=0},
        tickwidth=0pt,
        yticklabel style={
        /pgf/number format/fixed,
        /pgf/number format/precision=5
        },
        scaled y ticks=false,
        every axis title/.append style={at={(0.75, 0.0)}}, %
        legend image post style={scale=0.5}
	    ]
    \nextgroupplot[
		legend style = {
		  font=\scriptsize ,
          draw=none, 
          fill=none,
          column sep = 2pt, 
          /tikz/every even column/.append style={column sep=3mm},
          legend columns = -1, 
          legend to name = grouplegend},
		title= Normal users, 
		ylabel= average immediate reward,
		]
		
    \addplot[matisse, very thick] table [x=x, y=y, col sep=comma] {./figs/1_normal_smooth9.csv};  \addlegendentry{$p=1$}
    \addplot[flamingo, very thick] table [x=x, y=y, col sep=comma] {./figs/o2_normal_smooth9.csv}; \addlegendentry{$p=1/2$}
    \addplot[summer_sky, very thick] table [x=x, y=y, col sep=comma] {./figs/o4_normal_smooth9.csv}; \addlegendentry{$p=1/4$}
    \addplot[studio, very thick] table [x=x, y=y, col sep=comma] {./figs/o8_normal_smooth9.csv}; \addlegendentry{$p=1/8$}
    \addplot[sun_shade, very thick] table [x=x, y=y, col sep=comma] {./figs/o16_normal_smooth9.csv}; \addlegendentry{$p=1/16$}
    \addplot[black, very thick] table [x=x, y=y, col sep=comma] {./figs/s8_normal_smooth9.csv}; 
    
    \hfil
    \nextgroupplot[
    title=  Sensitive users,
    ylabel=  average immediate reward,
    ]
    \addplot[matisse, very thick] table [x=x, y=y, col sep=comma] {./figs/1_sen_smooth9.csv};
    \addplot[flamingo, very thick] table [x=x, y=y, col sep=comma] {./figs/o2_sen_smooth9.csv};
    \addplot[summer_sky, very thick] table [x=x, y=y, col sep=comma] {./figs/o4_sen_smooth9.csv};
    \addplot[studio, very thick] table [x=x, y=y, col sep=comma] {./figs/o8_sen_smooth9.csv};
    \addplot[sun_shade, very thick] table [x=x, y=y, col sep=comma] {./figs/o16_sen_smooth9.csv};
    \addplot[black, very thick] table [x=x, y=y, col sep=comma] {./figs/s8_sen_smooth9.csv}; 

\end{groupplot}
\node at ($(group c1r1) + (120pt, 100pt)$) {\ref{grouplegend}};
\end{tikzpicture}

\caption{Simulation process using ``oldest continuous'' strategy with different granularity $p$ on ML-100k. The results are smoothed using exponential moving average with smoothing factor 0.9 for clearer visualization.
The black line denotes the ``separate'' with $p{=}1/8$ on ML-100k truncated at 400 epochs. %
}
\label{fig:granularity}
\end{figure}

We study how data disclosure strategy affects users' decisions using three strategies with different degrees of freedom. %
The results are also reported in \cref{Rec_results}.

\begin{table*}
\renewcommand{\arraystretch}{1.03}
    \centering
    \caption{Results on different recommendation models with different platform mechanism.
    ``dis.\%'' denotes average user data disclosure percentage, ``NDCG'' means NDCG@100 (\%).
        All the results are averaged on the last 20 epochs.
        }
        \label{models_results}
    \begin{adjustbox}{max width=\textwidth}
        \begin{tabular}{l l r r r r r r r r} \toprule 
           &  & \multicolumn{8}{c}{ML-100k}  \\
             \cmidrule(lr){3-10}
           strategy & model & \multicolumn{2}{c}{non-sensitive}  & \multicolumn{2}{c}{normal} & \multicolumn{2}{c}{sensitive} & \multicolumn{2}{c}{all}  \\ \cmidrule(lr){3-4}  \cmidrule(lr){5-6} \cmidrule(lr){7-8}  \cmidrule(lr){9-10} 
            & & NDCG & dis.\% & NDCG & dis.\% & NDCG & dis.\% & NDCG & dis.\%  \\ 
            \midrule
           \multirow{3}{*}{\shortstack[l]{latest\\ continuous\\$p{=}1/16$}} 
           & NCF  &
          10.73 & 100 & 15.69 & 13.52  & 9.87 & 9.78 & 12.10& 39.82 \\
           & GRU4Rec  & 
          17.65 &  100 &  19.26 & \underline{16.22} & 12.01 & 8.82 & 16.31 &  \underline{40.45}\\
           & BiSA  & 
            \underline{18.04} &100 & \underline{21.88} & 15.61 & \textbf{14.55} &\underline{12.80} & \underline{18.16} & 40.44  \\
            \midrule
          \multirow{3}{*}{\shortstack[l]{oldest\\ continuous\\$p{=}1/16$}}
           & NCF  & 
           11.60 & 100 & 14.08 & 14.88&  10.11 & 7.80 & 11.93 &39.61 \\
           & GRU4Rec  & 
                 17.55 & 100 &  21.55 & 26.85 & 12.88 & 17.03 & 17.33 &  46.89  \\
            & BiSA  & 
           \textbf{18.05} &100 & \textbf{24.52} & \textbf{34.72} & \underline{13.35} & \textbf{17.73} & \textbf{18.64} & \textbf{49.85} \\
            \midrule
          \multirow{3}{*}{\shortstack[l]{separate\\$p{=}1/4$}}
           & NCF & %
          11.37 & 100 & 16.32 &17.57 & 9.06 &8.37 & 12.25 & 40.76  \\
           & GRU4Rec  &
           17.38 & 100 &  22.26 & 23.88  & 9.44 & 11.85 & 16.36 & 44.12  \\
            & BiSA  & 
          \underline{17.53} &100 & \underline{24.33} & \underline{26.40} & \underline{9.64} & \underline{12.42} & \underline{17.17} & \underline{45.42}\\
          
          \toprule 
          &  & \multicolumn{8}{c}{Yelp}  \\
             \cmidrule(lr){3-10}
            \multirow{3}{*}{\shortstack[l]{latest\\ continuous\\$p{=}1/16$}} 
       & NCF  &
      23.18 & 100 & 25.57 & 16.90 & 13.60 & 5.85  & 20.79 & \underline{41.41} \\
       & GRU4Rec  & 
     \underline{26.09} &  100 & 25.82 & \underline{17.22} & 15.62 & 6.08 &  22.33 &  41.10 \\
       & BiSA  & 
      25.35 & 100 & \underline{26.34} & 12.85& \textbf{19.48} &\underline{7.51} &  \textbf{23.72} &40.38 \\
        \midrule
      \multirow{3}{*}{\shortstack[l]{oldest\\ continuous\\$p{=}1/16$}}
       & NCF  & 
       23.21 & 100 & 21.13 & 15.97 & 10.22  &4.56 & 18.19 &40.71 \\
       & GRU4Rec  & 
      \underline{25.99} & 100 & 24.77 & 24.68 & 11.06 & 6.46 & 20.61 & 44.21    \\
        & BiSA  & 
      25.16 &100 & \textbf{31.53} & \underline{26.53} & \underline{14.07}  & \textbf{8.66} & \underline{23.59} & \textbf{45.18} \\
        \midrule
      \multirow{3}{*}{\shortstack[l]{separate\\$p{=}1/4$}}
       & NCF & %
      23.81 & 100 & 27.25 & 24.60 & 5.73 & 3.32 & 18.92 & 43.15 \\
       & GRU4Rec  &
      \textbf{ 26.48} & 100 & 29.97 & \textbf{27.73}  & 4.56  & 4.64 & 20.34 & \underline{44.61}  \\
        & BiSA  & 
      26.30 & 100& \underline{30.47} &26.32 & \underline{6.64} & \underline{5.02} & \underline{21.14} & 44.28 \\

            \bottomrule
        \end{tabular}
        \end{adjustbox}
    \end{table*}

It is easy to see that the flexible ``separate'' strategy is superior to other mechanisms within the same granularity. 
The ``separate'' strategy achieves better overall recommendation results with similar or even less disclosed data. %
One possible reason is that it enables users to freely disclose the data that benefits their recommendations.
In this way, users will discard those data that are not helpful for their recommendations, which is equivalent to data optimization by users.
It also explains why ``separate'' with $p{=}1/8$ outperforms ``all'' in ML-100K.
These results are consistent with the research in data minimization~\cite{chow2013differential,Wen:recsys18:Exploring,biega2020operationalizing}.

For the other two strategies, i.e., \czq{``latest continuous'' }and ``oldest continuous'' , the results show they perform not very consistently in different datasets.
This could be caused by the characteristics of different datasets.
It reminds us to design data disclosure mechanisms carefully according to the characteristics of data we deal with in real-world applications.

In summary, the platform mechanism affects both the possibility of a user to disclose and the volume of her/his disclosed data. 
Normally, a finer granularity and a more free data disclosure strategy can improve recommendation results for users and better protect users' privacy at the same time. 
However, the price is that a large action space leads to slow convergence on the optimal user policy. 
As shown in \cref{fig:granularity}, ``separate'' with $p{=}1/8$ (128 possible options) converges much slower than other mechanisms.
It indicates users might be hard to find their best policies under these fine-grained mechanisms.
Thus, we constraint users' possible choice number to 16 for all mechanisms for fair comparisons in the latter experiments.

\begin{table}
    \renewcommand{\arraystretch}{1.2}
    \centering
    \caption{%
    Results for different user group compositions on ML-100K. All the results are averaged on the last 20 epochs.
    }%
    \label{tab:user_composition}
    \begin{adjustbox}{max width=\linewidth}
    \begin{tabular}{l rrrr}
\toprule
\multicolumn{5}{c}{ \textbf{NDCG@100 (\%)}} \\ \toprule
Strategy & \#100\% Non-sen & \#100\%normal &  \#100 \% sensitive  &  \#1:1:1  \\ 
\midrule
\textbf{latest $p=1/16$} &  19.45 & 16.64  &   10.50  & 16.31    \\ 
\textbf{oldest $p=1/16$} &   19.45 &  18.76   &  11.87  & 17.33  \\
\textbf{separate $p=1/4$} &   19.45  & 19.28 & 9.27   &  16.36   \\
\bottomrule %
\multicolumn{5}{c}{\textbf{statistics of data disclosure (\% of disclosed data (avg. \# user))}} \\ \toprule
Strategy & \#100\% Non-sen & \#100\% normal &  \#100 \% sensitive  &  \#1:1:1  \\ 
\midrule
\textbf{latest $p=1/16$} & 100 (637)&  15.70 (429.7)&   11.66  (215.5) &  70.96 (449.9)    \\ 
\textbf{oldest $p=1/16$} &   100 (637) &  24.55 (378.0)   &  15.86 (201.3)  & 66.88 (411.9)    \\
\textbf{separate $p=1/4$} & 100 (637) & 25.36 (462.1) &  15.11 (163.4)  &  68.13 (422.2)  \\

\bottomrule
\end{tabular}
\end{adjustbox}
\end{table}

\subsection{Study 2: Impact of Recommender}

This study answers whether a better model (BiSA) or a worse model (NCF) will attract users to disclose more data or not. 
\cref{models_results} reports the results on three different data disclosure strategies with optimal granularity $p=1/16$ except the $p=1/4$ on ``separate'' for keeping the same number of data disclosing choices. 
It can be observed that:

\begin{enumerate}
    \item The results show that a more powerful model BiSA can attract sensitive users to disclose more data by improving the recommendation results for all platform mechanisms on all datasets.
    \item Though a better model usually can incentive users to disclose data, the total volume of data disclosed by normal users is not always increased. 
One reason is that marginal recommendation utility by disclosing more data may decrease on a better model considering the model already predicted precisely based on the disclosed data.
This phenomenon is prominent on the BiSA in ``latest continuous'' considering a better sequential model may rely less on older behavior data~\cite{kang2018self,Sun:cikm19:BERT4Rec}.
\end{enumerate}

In summary, 
the \cref{models_results} results suggest that the platform may pay more attention towards mechanism optimization while always stick to a better recommender system.

\subsection{Ablation Study}
Here, we study the impacts of compositions of user groups and the privacy sensitive hyper-parameter $w_i$. %
Due to page limitation, we only report the results based on GRU4Rec on ML-100k.

\subsubsection{User Group Compositions}

In this subsection, we adjust the user group composition where each user in the dataset is non-sensitive/normal/sensitive.
The average percentage of disclosed data and recommendation results are reported in \cref{tab:user_composition}.
The platform can get higher revenues when users are less concerned about their privacy risks.
Moreover, user group compositions play a more critical role in the platform's revenue than the data disclosure mechanisms.
This encourages the platform to take more actions on privacy protection to prevent users from becoming sensitive.

\subsubsection{Privacy Sensitive}

\pgfplotsset{
axis background/.style={fill=gallery},
grid=both,
  xtick pos=left,
  ytick pos=left,
  tick style={
    major grid style={style=white,line width=1pt},
    minor grid style=gallery,
    draw=none
    },
  minor tick num=1,
  ymajorgrids,
	major grid style={draw=white},
	y axis line style={opacity=0},
	tickwidth=0pt,
}

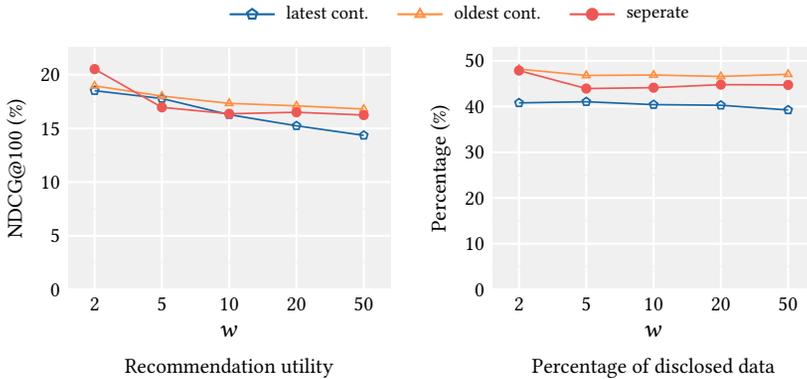
\begin{figure}
\centering
    \begin{tikzpicture}[scale=0.8]

    \begin{groupplot}[
	    group style={group size=2 by 1,
	        horizontal sep = 48pt,
	        },
	    width=0.5\textwidth,
	    height=0.4\textwidth,
        ymin = 0, 
	    xlabel= \Large $w$,
        xticklabels={2, 5, 10, 20, 50},
        xtick={1,2,3,4,5},
        ymajorgrids,
        major grid style={draw=white},
        y axis line style={opacity=0},
        tickwidth=0pt,
        yticklabel style={
        /pgf/number format/fixed,
        /pgf/number format/precision=5
        },
        scaled y ticks=false,
        every axis title/.append style={at={(0.5,-0.45)}}, %
	    ]
	    
	    \nextgroupplot[
		legend style = {
		  font=\scriptsize ,
          draw=none, 
          fill=none,
          column sep = 2pt, 
          /tikz/every even column/.append style={column sep=3mm},
          legend columns = -1, 
          legend to name = sensitivity},
		title= Recommendation utility, 
		ylabel= NDCG@100 (\%),
		ytick={0, 5, 10, 15, 20, 25},
		]
		
		\addplot[thick,color=matisse,mark=pentagon] coordinates {
          (1, 18.51)
          (2, 17.78)
          (3, 16.28)
          (4, 15.24)
          (5, 14.35)
        }; \addlegendentry{latest cont.}
         \addplot[thick,color=sun_shade,mark=triangle] coordinates {
          (1, 18.96)
          (2, 18.01)
          (3, 17.33)
          (4, 17.1)
          (5, 16.8)
        }; \addlegendentry{oldest cont.}
        \addplot[thick,color=flamingo,mark=*] coordinates {
          (1, 20.52)
          (2, 16.96)
          (3, 16.36)
          (4, 16.50)
          (5, 16.24)
        };\addlegendentry{seperate}
    \hfil
     \nextgroupplot[
        title=  Percentage of disclosed data,
        ylabel=  Percentage (\%),
        ytick={0, 10, 20, 30, 40, 50},]
     \addplot[thick,color=matisse, mark=pentagon] coordinates {
          (1, 40.8)
          (2, 41.05)
          (3, 40.42)
          (4, 40.28)
          (5, 39.25)
        }; 
         \addplot[thick,color=sun_shade,mark=triangle] coordinates {
          (1, 48.19)
          (2, 46.8)
          (3, 46.89)
          (4, 46.58)
          (5, 47.05)
        }; 
        \addplot[thick,color=flamingo,mark=*] coordinates {
          (1, 47.87)
          (2, 43.93)
          (3, 44.13)
          (4, 44.79)
          (5, 44.73)
        };
        
    \end{groupplot}
\node at ($(group c1r1) + (110pt,72pt)$) {\ref{sensitivity}};
\end{tikzpicture}

    \caption{
    Averaged results of different $w$ on ML-100K.
    }
    \label{fig:lambda}
\end{figure}

We report the effects of the hyper-parameter $w_i$  in \cref{fig:lambda}.
The results show that all mechanisms perform quite stable on both recommendation results and data disclosure with different hyper-parameter $w$, especially when $w>5$.
The reason for these results is that a user will barely disclose their data unless she/he observes a significant improvement on the recommendation results when her/his privacy sensitivity is very high (e.g., $w > 5$).
This also demonstrates that the conclusions of our previous experiments are stable.

\subsection{Summarized Insights}
\czq{
In this subsection, we summarize several mentioned insights as below.
}
\begin{enumerate} [itemsep=4pt]
\item \czq{We derive a negative attitude towards the  ``all or nothing'' binary mechanism due to its worst performances compared to other proposed mechanisms.}
\item \czq{The platform mechanism affects both the possibility of a user to disclose and the volume of her/his disclosed data. 
A finer granularity mechanism will normally attract more users to disclose while the volume of her/his disclosed data and the recommendation performances are not always monotonically increasing. }
\item \czq{Though a better model usually can incentive users to disclose data, the total volume of data disclosed by normal users is not always increased due to the marginal effects, which suggests the platform may pay more attention towards mechanism optimization based on a specific recommender system rather than always stick to the recommender system optimization.}

\end{enumerate}

\section{Related Work}
In this section, we will review previous works which are highly related to ours in the three fields, i.e., recommender systems, privacy research in recommender systems, and simulation.

\subsection{Recommendation Systems} 
Recommender systems play an essential role in today's web service platforms, e.g., e-commerce~\cite{Linden:IC03:Amazon,xie21explore} and social media~\cite{Covington:recsys16:Deep,Ying:kdd18:Graph}, since they provide a personalized and convenient tool for every user to alleviate the information overload problem or explore serendipity things.
Besides the attention of industry, recommender systems have also become the most active direction in information retrieval research~\cite{Zhang:csur19:Deep,Quadrana:csur19:Sequence,Wu:Graph}.

Early works on recommender systems mainly model the users' interests statically as collaborative filtering (CF) task with implicit feedback.
Early representative works include item-base CF algorithms~\cite{Sarwar:www01:Item,Linden:IC03:Amazon} and matrix factorization (MF)~\cite{Mnih:nips08:Probabilistic,Koren:Computer09:Matrix}
Recently, deep learning has also revolutionized collaborative filtering.
One line of research seeks to improve the CF models with the representation learned from auxiliary information, e.g., text~\cite{Wang:kdd15:Collaborative} and images ~\cite{Wang:www17:What} using deep learning models.
While more mainstream way is to take the place of conventional CF models with more powerful neural models, like neural collaborative filtering (NCF)~\cite{He:www17:Neural} and graph neural network based recommendation models~\cite{Ying:kdd18:Graph,He:sigir20:LightGCN}.

In recent years, sequential recommendation has become another mainstream task in recommender systems since it can better capture users' dynamic interests from their historical behaviors~\cite{Quadrana:csur19:Sequence}.
Sequential recommendation has also experienced the development process from traditional markov chain based models~\cite{Shani:kmlr05:MDP,Rendle:www10:Factorizing} to neural sequential models, e.g., GRU4Rec~\cite{Hidasi:ICLR2016:gru4rec,hidasi2018recurrent} and self-attention models~\cite{kang2018self,Sun:cikm19:BERT4Rec}.
Considering that sequential recommendation has become the mainstream in real-world applications~\cite{lv2019sdm,Li:cikm19:Multi}, we study the proposed task with sequential models in this paper.

\subsection{Privacy in Recommender Systems}
The research about privacy concerns in recommender systems can be classified into two categories: privacy-preserving recommendation modeling and decision making in privacy.

\textbf{Privacy-preserving recommendation modeling} mainly aims to protect user's sensitive information from being leaked by designing specific models.
An emerging paradigm is to use federated learning to train recommender systems without uploading users' data to the central server
\cite{Qi:emnlp20:Privacy,Muhammad:kdd20:FedFast,Lin:sigir20:Meta,wang:vldbj2021:fast}.
Federated learning dramatically enhances user privacy since user data never leaves their devices.
However, recent works have shown that federated learning can unintentionally leak information through gradients~\cite{Zhu:nips19:Deep,li2019privacy} and is also vulnerable to attacks like membership inference attacks~\cite{melis2019exploiting,nasr2019comprehensive}.
To address such issues, differential privacy~\cite{Dwork:Algorithmic}, a powerful mathematic framework for privacy, has been employed to guarantee user privacy in the procedure of recommender systems~\cite{McSherry:kdd09:Differentially,Berlioz:recsys15:Applying,shin2018privacy,Gao:sigir20:DPLCF}.
The basic idea of this paradigm is to add random noise into the recommender system to prevent information leakage.
As a promising framework, one limitation of differential privacy is that it usually decreases performance~\cite{Domingo:cacm21:Limits}.

\textbf{Decision making in privacy} from other disciplines, e.g., economic~\cite{lin2019valuing}, management sciences~\cite{Culnan:os99:Information}, and human–computer interaction~\cite{Knijnenburg:recsys12:Inspectability,Knijnenburg:tiis13:Making}, mainly focus on studying the problem like where privacy concerns come from and how to mitigate them.
They mainly study the procedure of user's decision making about information disclosure using the \textit{privacy calculus theory}~\cite{Laufer:si77:Privacy,Culnan:os99:Information}, which views privacy as an economic commodity.
It is to say that the user decides to disclose his/her information by weighing the anticipated risks of disclosing personal information against the perceived utility.
Numerous works have studied the factors that influenced the user's decision using questionnaires or mock-up applications~\cite{Knijnenburg:recsys12:Inspectability,Knijnenburg:tiis13:Making,Chen:CHI18:This,Zhang:hcs19:Proactive}.
Multiple studies highlight that ``control'' is a key factor in decision making about privacy, and providing control over the recommendation process to users can reduce their privacy concerns~\cite{Zhang2014-oa,Chen:CHI18:This}.
Going a step further, in this paper, we give users not only control over whether or not to disclose data, but also control over which data to disclose.
Then we investigate the consequences caused by this novel setting, including how users make choices and how different platform mechanisms and recommendation models perform.

Another close work to ours is \cite{Xin:nips14:Controlling} that studies a recommendation task where a small set of ``public'' users who disclose all their ratings (large amount) and a large set of ``private'' users refuse to disclose their data.
Our work differs from \cite{Xin:nips14:Controlling} in the following aspects:
\begin {enumerate*} [label=\roman*\upshape)]
\item Most importantly, as explained in the introduction and last paragraph, our goal is not the performances of the recommender systems;
\item We provide users with more fine-grained control over their data; 
\item our task is built on implicit feedback, which is the mainstream of the real-world applications.
\end {enumerate*}

\subsection{Simulation}

Recent years have witnessed the wide applications of the simulation techniques among various scenarios, e.g., recommendation \cite{jannach2015recommenders,lucherini2021t,chaney2018algorithmic,yao2017beyond}, autopilot \cite{osinski2020simulation}, traffic scheduling \cite{chu2019multi,abdoos2020cooperative} and robotic \cite{rao2020rl}.
The primary reason to utilize simulation is that straightforwardly conducting experiments in the real world may remain too expensive~\cite{virtual_taobao} and risky \cite{osinski2020simulation}.
Besides, the solutions derived from simulations can be transferred to solve the real-world problems~\cite{virtual_taobao,tobin2017domain}.

In the research areas of recommender systems, it is of great significance to utilize the carefully designed simulation environments to efficiently evaluate recommendation policy~\cite{virtual_taobao} or draw insightful conclusions for specific studies such as societal impact analysis~\cite{chaney2018algorithmic}.
\citet{ie2019reinforcement} builds upon a simulation environment for slate-based recommender systems, which facilitates the recommendation policy evaluation. 
\citet{virtual_taobao} proposes to utilize the historical user behavior data to train the simulator and verifies that the policies trained in the simulator achieve superior online performance.
A series of works \cite{lucherini2021t,chaney2018algorithmic,yao2017beyond} utilize simulation environments to get in touch with user society impacts over recommender systems such as fairness and societal biases. 

There also exist a series of works to simulate user decisions to maximize their personalized utilities~\cite{jiang2017information,samadi2012advanced,kallstrom2019tunable}.
Typically, the user is modeled as a rational agent whose policy can be learned following a trial-and-error schema such as RL-based algorithms \cite{katehakis1987multi,kallstrom2019tunable}.
In this work, each user is modeled as a rational agent to optimize his (her) unique data disclosure policy under a designed platform mechanism. The efficiency of recommender systems is also evaluated within such a simulation environment.

\section{Conclusions and Future Work}

This paper proposes a privacy aware recommendation framework based on privacy calculus theory to study what will happen if the platform gives users control over their data.
To avoid the great cost in online experiments, we propose to use reinforcement learning to simulate the users' privacy decision making under different platform mechanisms and recommendation models on public benchmark datasets.
The results show a well-designed data disclosure mechanism can perform much better than the popular ``all or nothing'' binary mechanism.
Our work provides some insights to improve current rough solutions in privacy protection regulations, e.g., opt-in under GDPR and opt-out under CCPA.

This paper only takes the first step in studying users' privacy decision making under different platform mechanisms, and several directions remain to be explored.
First, a more complex and accurate privacy cost function can help us better understand users' privacy decision making.
In this work we have modeled different users with their individual privacy sensitivity weights, and one may modify the privacy cost function on the effects of the users' trust towards the platform in the future.
Second, more sophisticated platform mechanisms are also worth exploring.
Recent mechanism design works also turn to the perspectives of deep neural network based mechanism designs, which can be explored with our proposed framework.
Last but not least, deploying online experiments and analyzing users' decisions in real-world can facilitate further researches.

\clearpage

\bibliographystyle{ACM-Reference-Format}
\bibliography{sample-base}

\end{document}